\documentclass[sigconf]{acmart}
\usepackage{amsmath}
\usepackage{bm}
\usepackage{multicol}

\renewcommand{\baselinestretch}{0.985}

\AtBeginDocument{%
  \providecommand\BibTeX{{%
    \normalfont B\kern-0.5em{\scshape i\kern-0.25em b}\kern-0.8em\TeX}}}

\setcopyright{rightsretained}
\begin{document}
\fancyhead{}
\copyrightyear{2021}
\acmYear{2021}
\acmConference[WSDM '21]{Proceedings of the Fourteenth ACM International Conference on Web Search and Data Mining}{March 8--12, 2021}{Virtual Event, Israel}
\acmBooktitle{Proceedings of the Fourteenth ACM International Conference on Web Search and Data Mining (WSDM '21), March 8--12, 2021, Virtual Event, Israel}\acmDOI{10.1145/3437963.3441798}
\acmISBN{978-1-4503-8297-7/21/03}

\title{Non-Clicks Mean Irrelevant? Propensity Ratio Scoring As a Correction}


\author{Nan Wang$^1$, Zhen Qin$^2$, Xuanhui Wang$^2$, Hongning Wang$^1$}
\affiliation
{
    \institution{$^{1}$University of Virginia, Charlottesville, VA}
 }
\affiliation
{
    \institution{$^{2}$Google Research, Mountain View, CA}
 }
\email{nw6a@virginia.edu,{zhenqin, xuanhui}@google.com,hw5x@virginia.edu}




\renewcommand{\shortauthors}{Nan Wang, Zhen Qin, Xuanhui Wang, Hongning Wang.}

\begin{abstract}
Recent advances in unbiased learning to rank (LTR) count on Inverse Propensity Scoring (IPS) to eliminate bias in implicit feedback. Though theoretically sound in correcting the bias introduced by treating clicked documents as relevant, IPS ignores the bias caused by (implicitly) treating non-clicked ones as irrelevant. In this work, we first rigorously prove that such use of click data leads to unnecessary pairwise comparisons between relevant documents, which prevent unbiased ranker optimization. 
Based on the proof, we derive a simple yet well justified new weighting scheme, called Propensity Ratio Scoring (PRS), which provides treatments on both clicks and non-clicks. Besides correcting the bias in clicks, PRS avoids relevant-relevant document comparisons in LTR training and enjoys a lower variability. Our extensive empirical evaluations confirm that PRS ensures a more effective use of click data and improved performance in both synthetic data from a set of LTR benchmarks, as well as in the real-world large-scale data from GMail search.

\end{abstract}



\keywords{Unbiased learning to rank, implicit feedback}

\maketitle

\section{Introduction}
 
Implicit feedback from users, such as clicks, provides an abundant resource of relevance signals for learning to rank (LTR) \cite{Joachims2002OSE}. But such data is also notoriously biased for various reasons, e.g., the most notable position bias \cite{Joachims2017interpretclick}, which is known to distort LTR training if not properly handled \cite{Joachims2007Eval, Joachims2017interpretclick, yue2010beyondposition}.

Inverse Propensity Scoring (IPS)  \cite{joachims2017ips} has emerged as a mainstream solution to debias implicit feedback for LTR.
It provides an unbiased estimate of the ranking metric of interest, such as Average Relevance Position (ARP, termed as $rank$ in \cite{joachims2017ips}), by reweighing the clicked documents. 
However, the unbiasedness of IPS is only maintained in the estimation of the ranking metrics, rather than in the actual ranker optimization under such metrics. A serious gap is introduced when one evaluates those ranking metrics using click data for ranker optimization. Specifically, due to the non-continuous nature of most ranking metrics, the optimization has to be performed on induced loss of those metrics in practice. 
As shown in~\cite{Wang2018Lambdaloss}, most popular ranking metrics, such as ARP and NDCG, can be decomposed into pairwise comparisons. Thus pairwise loss is introduced for continuous approximation and ranker optimization under those metrics. For example, in a ranked list, each clicked document is compared against all others to compute the hinge loss for ARP as in Propensity SVM-Rank \cite{joachims2017ips}, or against only non-clicked ones to compute the lambda loss for NDCG as in Unbiased LambdaMART \cite{hu2019unbiasedlambda}. 
The mapping from a ranking metric estimator to its induced loss by clicks opens the door to the detrimental deficiency of IPS-based unbiased LTR, which is serious but largely ignored. 

In this work, we rigorously prove the gap between the IPS estimators of the ranking metrics and the practical ranker optimization on the induced losses, in terms of unbiasedness, for the first time. In particular, we show that comparisons between pairs of the same relevance label only contribute a constant term to the evaluation of ranking metrics, regardless of their ranked positions or ranking scores. Thus one should avoid counting loss on such pairs. 
Because IPS \cite{joachims2017ips, Agarwal2019GFC, ai2018DLA} can only correct bias in using clicked documents to measure relevance, but non-click does not necessarily stand for irrelevance (it can be relevant but not observed), existing IPS-based solutions inevitably count loss on relevant-relevant document comparisons. This introduces an irreducible gap between the ranking metric one expects to optimize and the actual loss one ends up with. 


\begin{figure}[t]
    \centering
    \includegraphics[width=0.96\linewidth]{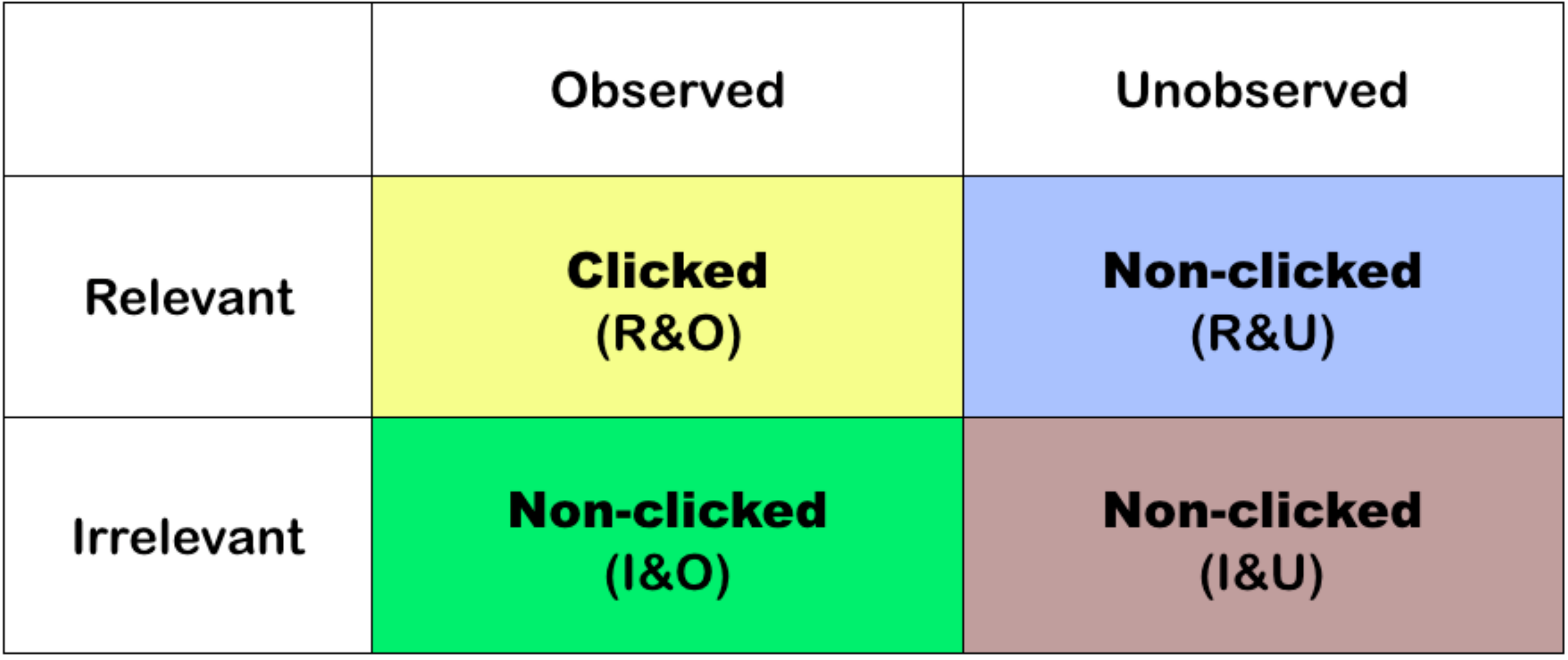}
    \caption{Decomposition of click data. We use `R', `I' to denote relevant or irrelevant; and `O', `U' for observed or unobserved. Only documents in the R\&O part are clicked.}
    \label{fig:clickdata}
\vspace{-1.5em}    
\end{figure}

To illustrate the problem in a more intuitive way, we decompose click data as shown in Figure \ref{fig:clickdata}. In a noise-free setting, click happens if and only if a relevant document is observed (i.e., R\&O), while all others are recorded as non-clicks (i.e., R\&U, I\&O and I\&U in the figure). This leads to an \emph{asymmetric} relation between clicked and non-clicked documents. 
When using click data for LTR training, IPS can solely correct bias in the clicked part, so that the total loss can be extended from clicked documents to all relevant documents (R\&O+R\&U) in terms of expectation. Meanwhile, relevant and unobserved documents (R\&U) are also in the non-clicked documents, and are unavoidably used as negative examples in pairwise comparisons in the induced loss for LTR \cite{joachims2017ips,hu2019unbiasedlambda}.
It leaves the learnt ranker still biased for relevant but non-clicked documents. 
This deficiency is already reflected in recent empirical studies \cite{Jagerman2019modelorintervene}: only when there is little bias, can the utility of IPS be observed. Because strong bias conceals lots of relevant documents in the non-clicked part, an increased gap is introduced to the loss function and thus distorts the learnt ranker more seriously. 
No existing work realizes this deficiency of IPS in unbiased LTR yet. A recent work named Unbiased LambdaMART reweighs non-clicked documents with IPS for debiasing purpose \cite{hu2019unbiasedlambda}. However, as it does not analyze the exact bias in non-clicked documents and still restricts itself in the IPS framework, a rigid assumption has to be imposed that the probability of non-click is proportional to the probability of being irrelevant. This unfortunately is against most click modeling assumptions. 

To eliminate the gap, we propose a simple yet effective solution called Propensity Ratio Scoring (PRS) for unbiased LTR with pairwise comparisons. 
We devise statistical treatments on both clicked and non-clicked documents when forming pairwise loss for metric optimization. 
Both theoretical and empirical justifications of PRS are provided step by step. More importantly, we show that PRS ensures a more effective use of click data and reduced variability compared to IPS, without any additional requirement on the infrastructure or data collection. 
We conduct extensive empirical studies based on a set of LTR benchmark datasets to demonstrate the correctness and advantage of PRS. To show its utility in real-world industrial setting, we also evaluate PRS on the large-scale GMail search data, which further confirms its practical significance.

\section{Related Work}
Click data is a vital resource for LTR model training in modern retrieval systems. But the intrinsic bias, especially the position bias, greatly limits its effective use \cite{Joachims2007Eval,yue2010beyondposition}. Numerous click models \cite{chuklin2015click, Craswell2008clickpositionbias, Chapelle2009dynamicbayesian, Dupret2008UBM, Wang2013contentaware} have been proposed to model the bias in users' click behaviors, so as to extract true relevance labels. But they require repeated observations for reliable relevance inference. Researchers also adopt randomization techniques to eliminate bias \cite{Chapelle2012interleave,yue2009interactively,Swaminathan2015batchlearning} when collecting clicks. 
Though assumption free, randomization degrades the ranking quality during data collection and inevitably hurts user experience. 

Recent efforts focus on unbiased LTR directly from biased click data. 
The key idea is to obtain an unbiased estimator of a ranking metric of interest by leveraging statistical patterns embedded in the click data. 
Inverse propensity scoring (IPS) \cite{joachims2017ips,Agarwal2019GFC} is a mainstream solution, which reweighs observational clicks for debiasing. 
However, IPS is solely applicable to debias clicked documents, and its utility largely depend on whether most relevant documents can be revealed by the clicks. When there is increasing bias or noise, more relevant documents are buried in non-clicked documents, which IPS cannot handle. To make things worse, IPS-based methods treat non-clicked but relevant documents as negative samples for ranker training \cite{hu2019unbiasedlambda}. 
This directly leads to its poor performance in practice \cite{Jagerman2019modelorintervene}, and we will theoretically prove the cause of this deficiency. 

There exist various realizations of the IPS framework for unbiased LTR \cite{Agarwal2019GFC,ai2018DLA, hu2019unbiasedlambda}. Though applied on different model structures, they suffer from the same issues of IPS. Among them, Hu el al. \cite{hu2019unbiasedlambda} applied IPS to both clicked and non-clicked documents for debiasing. However, this solution ignores the fact that the bias in clicked and non-clicked documents is \emph{asymmetric}, and simply assumes that non-click probability is proportional to the irrelevance probability to apply IPS. In contrast, we carefully analyze the bias introduced in non-clicked documents and derive a justified solution that accounts for the use of both clicks and non-clicks. 
\section{Diagnosis of IPS in Practical Use}

In this section, we first present the general theory of IPS for unbiased ranking metric estimation. 
Then based on an in-depth discussion of IPS in ranker optimization, we elaborately investigate and disclose the deficiency of IPS in solving the problem. 

\subsection{IPS for Unbiased Metrics Estimation}
\label{sec:unbiaLTR}

Without loss of generality, we are given a set of i.i.d. queries $\bm{Q}$, where each query $q$ is associated with a list of candidate ranking documents $\{x_i\}^{|q|}_{i=1}$.
The goal of LTR is to optimize a ranker $\pi$ on $\bm{Q}$ under ranking metrics of interest (e.g., ARP, NDCG, etc). Formally, the empirical risk of $\pi$ incurred on $\bm{Q}$ can be obtained as, 
\begin{equation}
    R(\pi|Q) = \sum_{q\in\bm Q}R(\pi|q)=\sum_{q\in\bm Q}\sum_{x_i:r_q(x_i)=1}\Delta(x_i|\pi_q)
    \label{eq_empirical_risk}
\end{equation}
where $r_q(x_i)$ is the ground-truth relevance label of document $x_i$ to query $q$ (assuming binary relevance for simplicity), $\pi_q$ denotes the ranking of documents under query $q$, and $\Delta(x_i|\pi_q)$ measures the  contribution from a relevant document $x_i$ to the ranking metric, e.g., the $rank$ of $x_i$ in ARP or the discounted gain of $x_i$ in NDCG. 

Unlike the full-information setting where the relevance labels of all documents in $\pi_q$ are known, $r_q$ is only partially observed in the implicit click feedback, due to various click biases \cite{richardson2007predicting}. 
Simply using clicked documents to realize Eq \eqref{eq_empirical_risk} leads to a biased estimate of the ranking metric. 

IPS addresses the bias issue via weighing each clicked document in Eq \eqref{eq_empirical_risk} by the inverse of its observation propensity in the logged ranking $\tilde\pi_q$ \cite{joachims2017ips}. To be more specific, denote $o_q$ as a binary vector indicating whether the documents' relevance labels in $r_q$ are observed in $\tilde\pi_q$. For each element of $o_q$, the marginal probability $P(o_q(x_i)=1|\tilde\pi_q)$ is referred to as the observation propensity of document $x_i$. Consider the deterministic noise-free setting in \cite{joachims2017ips}, 
a document is clicked if and only if it is examined (thus observed) and relevant, i.e., $c_q(x_i)\Leftrightarrow o_q(x_i)\wedge r_q(x_i)$. We can then get an unbiased estimate of $R(\pi_q|q)$ for any new ranking $\pi_q$  via IPS \cite{Imbens2015causal,Paul1983central},
\begin{align}
\label{eq:IPS}
    R_{IPS}(\pi_q|q,\tilde\pi_q,o_q) &= \sum_{x_i:c_q(x_i)=1}\frac{\Delta(x_i|\pi_q)}{P(o_q(x_i)=1|\tilde\pi_q)} \\\nonumber
    &= \sum_{x_i:\,o_q(x_i)=1\, \wedge\, r_q(x_i)=1}\frac{\Delta(x_i|\pi_q)}{P(o_q(x_i)=1|\tilde\pi_q)}
\end{align}
where we introduce another binary vector $c_q$ to denote whether a document is clicked under $q$ in $\tilde\pi_q$. 

Although Eq \eqref{eq:IPS} has been proved to be an unbiased estimate of $R(\pi_q|q)$ for any new ranking $\pi_q$: $E_{o_q}\big[R_{IPS}(\pi_q|q,\tilde\pi_q,o_q)\big]=R(\pi_q|q)$ \cite{joachims2017ips}, a direct optimization of Eq \eqref{eq:IPS} is intractable, due to the non-continuous nature of most ranking metrics. Approximations are thus necessary for optimization \cite{Agarwal2019GFC}, which however introduce a gap from the estimated metric to the induced loss. Next we rigorously examine and analyze the consequence caused by this gap.   


\subsection{Issues of IPS in Practical Ranker Optimization}
\label{sec:metric-optimization}

In Eq \eqref{eq:IPS}, it is important to realize that the individual contribution $\Delta(x_i|\pi_q)$ of a relevant document $x_i$ has to be obtained from its comparisons to other documents in the ranking list $\pi_q$, as the ranking position of $x_i$ in $r_q$ depends on how many other documents are predicted to be more relevant than it by the ranker. As shown in \cite{Wang2018Lambdaloss}, this is achieved via pairwise comparisons for a set of popular ranking metrics in practice. For instance, to get the $rank$ of a relevant document $x_i$ in $\pi_q$, we need to compare its predicted relevance $\hat r_q(x_i)$ to all other documents' $\hat r_q(x_j)$. One key property in such a pairwise formulation is that only comparisons between two documents of \emph{different} ground-truth relevance labels have influence on the ranking metric evaluation and can thus contribute to the optimization. The comparisons between two documents of the \emph{same} relevance label will only form a constant term, agnostic to their positions or predicted relevance \cite{Wang2018Lambdaloss}. 

The gap thus emerges when using continuous approximations on the pairwise comparisons, such as logistic loss \cite{Arias2008logistic}, hinge loss \cite{Joachims2002OSE} or LambdaRank loss \cite{Burges2006lambdarank}: the induced loss on comparisons between same-labeled documents become an irreducible term in the total loss that distorts the optimization and leads to sub-optimal results. 
To theoretically and more explicitly illustrate the issue, we revisit the optimization of Average Relevance Position (ARP), which is analyzed in the first IPS-based LTR solution \cite{joachims2017ips} (but termed as $rank$). The same analysis can be easily applied to the optimization of other ranking metrics, such as NDCG \cite{Wang2018Lambdaloss}. 

Assume there are $n$ documents in $\pi_q$ and the rank of documents starts from 0. Use $r_q^i$ and $\hat r_q^i$ to replace $r_q(x_i)$ and $\hat r_q(x_i)$ for simplicity. We have the APR evaluated on $\pi_q$ as: 
\begin{align}
\label{eq:APR-loss}
    ARP
    =& \frac{1}{n}\sum_{i=1}^n rank(x_i|\pi_q)\cdot r_q^i
    = \frac{1}{2n}\sum_{i=1}^n\sum_{j=1}^n\Big(r_q^i\mathbb I_{\hat r_q^i<\hat r_q^j}+r_q^j\mathbb I_{\hat r_q^j<\hat r_q^i}\Big)\nonumber \\
    =&\frac{1}{n}\sum_{i=1}^n\sum_{j:r_q^j<r_q^i}|r_q^i-r_q^j|\mathbb I_{\hat r_q^i<\hat r_q^j}+C_1+C_2 \nonumber\\
    (*)\leq&\frac{1}{n}\sum_{i=1}^n\sum_{j:r_q^j<r_q^i}|r_q^i-r_q^j|\log\big(1+e^{-(\hat r_q^i-\hat r_q^j)}\big)+C_1+C_2
\end{align}
where $C_1=\frac{1}{2n}\sum_{i=1}^n\sum_{j:r_q^i=r_q^j}r_q^j$ and $C_2=\frac{1}{n}\sum_{i=1}^n\sum_{j:r_q^j<r_q^i}r_q^j$ ($C_2=0$ in binary case) are constants. As shown in the step $*$, we should only impose pairwise loss to bound the comparison indicator $\mathbb I_{\hat r_q^i<\hat r_q^j}$ on document pairs with different ground-truth relevance labels, for optimizing the metric (i.e., relevant-irrelevant pairs in this binary case). Comparisons from pairs with the same label form a constant $C_1$ that does not contribute to the optimization. 

More concretely, in $C_1$, a pair of documents with the same label $r$ count $r\mathbb{I}_{\hat r_q^i < \hat r_q^j} + r\mathbb{I}_{\hat r_q^j < \hat r_q^i} = r$ in the metric, independent of how they are ranked in $\pi_q$. If we introduce pairwise loss on such pairs, we are forcing the pair of documents to have the same predicted relevance values.
Using pairwise logistic loss as an example, the loss $r\cdot\big[\log\big(1+e^{-(\hat{r}_q^i - \hat{r}_q^j)}\big) + \log\big(1+e^{-(\hat{r}_q^j - \hat{r}_q^i)}\big)\big]$ is minimized only when $\hat{r}_q^i = \hat{r}_q^j$. In other words, the loss on a pair of documents with the same label impose unnecessary constraints that distort the optimization.  

Eq \eqref{eq:APR-loss} reveals the root cause to the issue of IPS in practical ranker optimization using implicit feedback. We are now prepared to present our general deficiency diagnosis of existing IPS-based LTR solutions. Consider a general pairwise loss $\delta(x_i,x_j|\pi_q)$ defined on two different documents $x_i$ and $x_j$ in $\pi_q$, which indicates how likely $x_i$ is more relevant than $x_j$. Following the IPS estimator in Eq~\eqref{eq:IPS}, the individual contribution of a relevant document $x_i$ to the ranking metric is thus upper bounded by the total pairwise loss of comparing $x_i$ to all other documents in $\pi_q$:
\begin{equation}
\label{eq:bound}
    \Delta(x_i|\pi_q) \leq \sum_{x_j\in\pi_q\wedge x_j\neq x_i} \delta(x_i,x_j|\pi_q)
\end{equation}
where we assume minimizing the loss will optimize the metric. 

In click data, where only the relevant and observed documents are indicated by clicks, the local loss on $q$ can be derived from Eq~\eqref{eq:IPS} with pairwise loss. There are currently two strategies for imposing the pairwise loss. The first one is to compare each clicked document to \emph{all} other documents as in Propensity SVM-Rank \cite{joachims2017ips} or its generalizations \cite{Agarwal2019GFC};
but in expectation, it contains loss on all relevant-relevant pairs. 
To avoid explicitly comparing two relevant documents indicated by clicks, the second strategy restricts the loss on comparisons between clicked and non-clicked ones \cite{rendle2009bpr,hu2019unbiasedlambda}:
\begin{align}
\label{eq:click-unclick}
    l_{IPS}(\pi_q|q,\tilde\pi_q,o_q) = \mathop{\sum_{x_i:c_q(x_i)=1}}\frac{\sum_{x_j:c_q(x_j)=0}\delta(x_i,x_j|\pi_q)}{P(o_q(x_i)=1|\tilde\pi_q)}
\end{align}
But as relevant and unobserved documents are also non-clicked, this strategy still cannot address the problem of including loss on relevant-relevant pairs that only distracts the optimization: 
\begin{align}
\label{eq:IPS-expectation}
    &E_{o_q}[l_{IPS}(\pi_q|q,\tilde\pi_q,o_q)] \\\nonumber
    =& \sum_{x_i:r_q(x_i)=1}E_{o_q}\bigg[\frac{o_q(x_i)\cdot\sum_{x_j:c_q(x_j)=0}\delta(x_i,x_j|\pi_q)}{P(o_q(x_i)=1|\tilde\pi_q)}\bigg] \\\nonumber
    =& \sum_{x_i:r_q(x_i)=1}\frac{P(o_q(x_i)=1|\tilde\pi_q)\cdot\sum_{x_j:c_q(x_j)=0}\delta(x_i,x_j|\pi_q)}{P(o_q(x_i)=1|\tilde\pi_q)} \\\nonumber
    =& \sum_{x_i:r_q(x_i)=1}\bigg(\sum_{x_j\in\pi_q:r_q(x_j)=0}+\mathop{\sum_{x_j:o_q(x_j)=0}}_{\wedge r_q(x_j) = 1 }\bigg)\delta(x_i,x_j|\pi_q)
\end{align}
In expectation, 
besides the pairwise loss on relevant-irrelevant pairs that we need, this strategy also includes pairs on relevant documents versus relevant but unobserved documents (the second sum in the bracket). The negative effect can also be perceived from an intuitive perspective: as the relevant but unobserved documents are mistakenly used as negative examples, it will degrade the new ranker's ability to recognize these missed relevant documents, making the ranker still unfavorably biased against them.

\begin{figure}[t]
    \centering
    \includegraphics[width=0.45\linewidth]{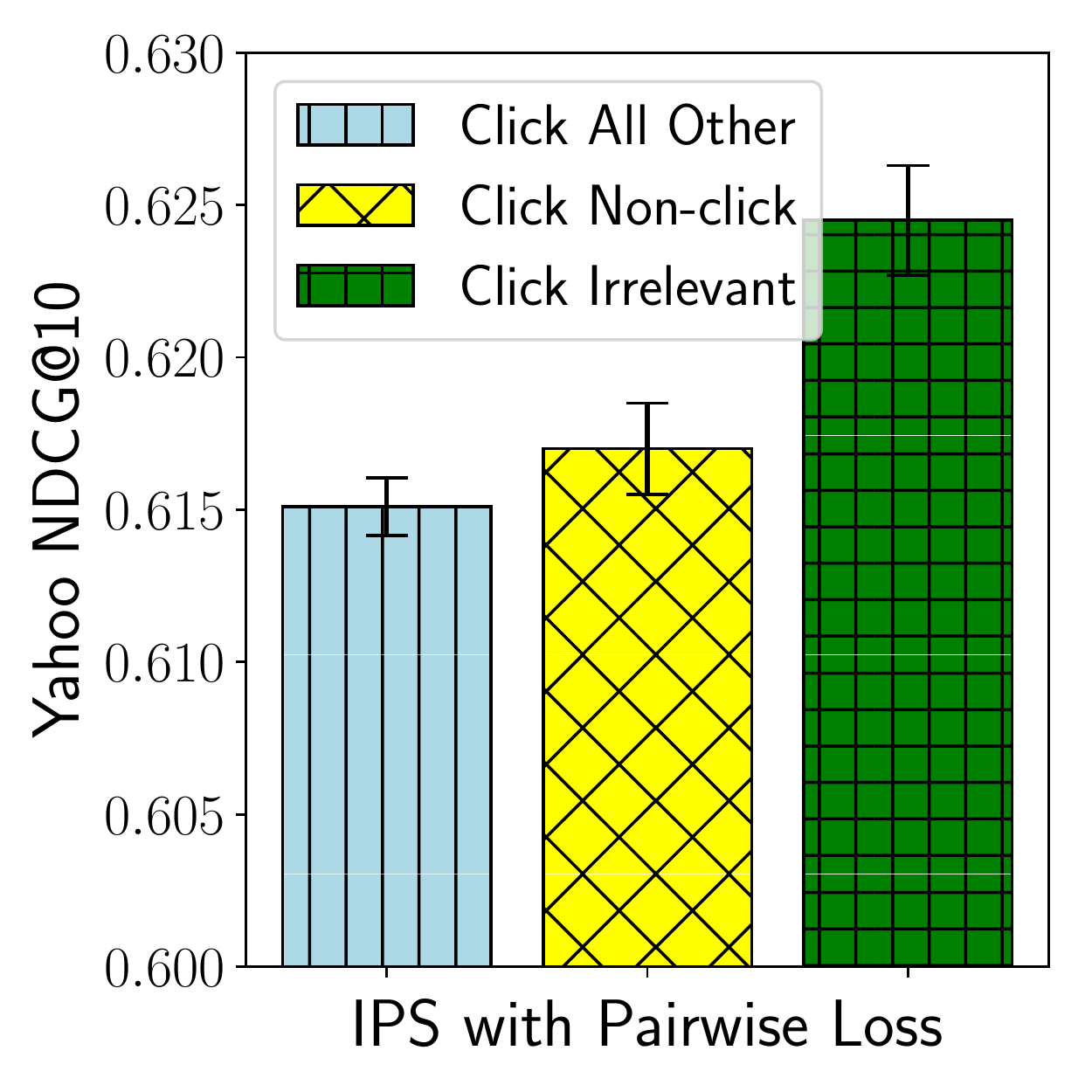}
    \includegraphics[width=0.45\linewidth]{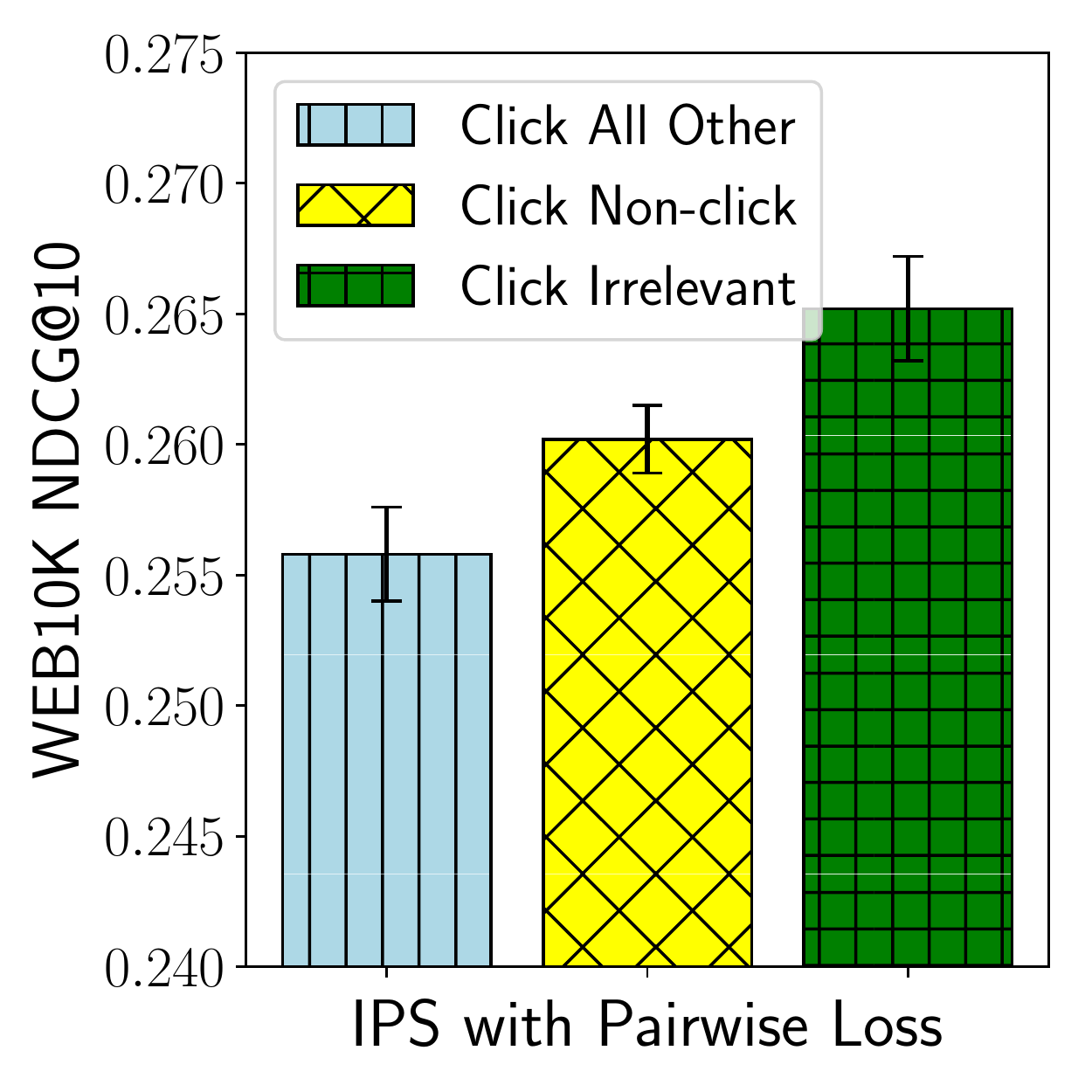}
    \vspace{-2mm}
    \caption{Empirical performance of applying IPS with pairwise loss. We compare the results of including pairwise loss on each clicked document against (1) all other documents; (2) non-clicked documents; (3) only irrelevant documents.}
    \label{fig:effect}
    \vspace{-4mm}
\end{figure}

In order to verify our theoretical analysis, in Figure \ref{fig:effect}, we demonstrate the empirical influence of including pairwise loss on relevant-relevant pairs in LTR model training with clicks. Specifically, we synthesize clicks on the Yahoo and Web10K LTR datasets with the propensity model ($\eta=1$) described in Section \ref{sec:exp}. To better illustrate the effect, we adopt the noise-free setting and sampled 128,000 clicks. To learn new rankers from the clicks, we apply IPS on the pairwise logistic loss, and include pairs from different comparison strategies. The NDCG@10 results are reported on fully labeled test sets in the two benchmarks accordingly. As clearly shown, including relevant-relevant document comparisons seriously hurt ranker optimization. 
Therefore, we should remove the relevant-relevant pairs to eliminate such adverse effects for LTR model learning, which will be the focus of next section. 

\section{Propensity Ratio Scoring}
\label{sec:PRS-forunbiasedLTR}
We have shown that to effectively tackle unbiased LTR in practice, we need to eliminate the relevant documents in non-clicked ones when counting loss, and thus keep the total loss unbiased for all relevant documents. 
In this section, we first derive a solution of identifying truly irrelevant documents from non-clicked ones, so as to avoid using relevant documents as negative examples to the best extent. Then we develop a holistic treatment on using click data for optimizing the ranking metric with pairwise comparisons. 

\subsection{Propensity-weighted Negative Samples}
\label{sec:PNS}
The insight of identifying truly irrelevant documents in non-clicked documents comes from the decomposition of click data in Figure \ref{fig:clickdata}. In a noise-free setting, if a document is observed (the first column in Figure \ref{fig:clickdata}), click is equivalent to relevance, i.e., $o_q(x_i)=1 \rightarrow [c_q(x_i)=r_q(x_i)]$. Consequently, if a document is observed and non-clicked, it is irrelevant, i.e., $[o_q(x_i)=1 \wedge c_q(x_i)=0] \rightarrow r_q(x_i)=0$
(the I\&O part in Figure \ref{fig:clickdata}). The problem of identifying truly irrelevant documents from non-clicked ones is thus reduced to finding the observed documents in the non-clicked ones. 


Note that for a non-clicked document $x_j$, we do not have its true observation $o_q(x_j)$. However, by weighting the loss on each non-clicked document with its conditional observation probability $P(o_q(x_j)=1|c_q(x)=0, \tilde\pi_q)$, we can restrict the loss on non-clicked documents to those that are non-clicked but observed, in expectation. But this conditional probability is not easy to estimate: As shown in Figure \ref{fig:clickdata}, this probability corresponds to the ratio of the $I\&O$ part in the non-clicked documents, which depends on both the position and relevance of the documents. Therefore, we need to estimate $P(o_q(x_j)=1|c_q(x)=0, \tilde\pi_q)$ on all positions under every single query. Without sufficient observations under the same query, the estimation quality can hardly be guaranteed. Instead, we propose to directly use the position-based observation propensity $P(o_q(x_j)=1|\tilde\pi_q)$ as an approximation for the purpose of reweighing the loss on each non-clicked document,
\begin{align}
\label{eq:PNS-expectation}
    &\sum_{x_j:c_q(x_j)=0}\Omega(x_j|\pi_q)\cdot P(o_q(x_j)=1|\tilde\pi_q) \\\nonumber
    =&\; \Big(\sum_{x_j:r_q(x_j)=0} + \mathop{\sum_{x_j:r_q(x_j)=1}}_{\wedge o_q(x_j)=0}\Big) \Omega(x_j|\pi_q)\cdot P(o_q(x_j)=1|\tilde\pi_q) \\\nonumber 
    =&\; E_{o_q}\Big[\mathop{\sum_{x_j:r_q(x_j)=0}}_{\wedge o_q(x_j)=1}\Omega(x_j|\pi_q)\Big] + \mathop{\sum_{x_j:r_q(x_j)=1}}_{\wedge o_q(x_j)=0}\Omega(x_j|\pi_q)\cdot P(o_q(x_j)=1|\tilde\pi_q) \,, 
\end{align}
where $\Omega(\cdot)$ represents any general loss on a non-clicked document $x_j$, including the pairwise loss used in Section \ref{sec:metric-optimization}. Due to the approximation, the second term of Eq \eqref{eq:PNS-expectation} still contains relevant documents; but $P(o_q(x_j)=1|\tilde\pi_q)$ for relevant but non-clicked documents is expected to be small. 
For example, under a position-based examination model \cite{joachims2017ips}, the examination probability shrinks fast over positions; and thus the impact from relevant but non-clicked documents can be quickly eliminated when moving down the ranked list.
We refer to this weighting on non-cliked documents as the Propensity-weighted Negative Samples (PNS). 

\subsection{Debiasing Pairwise Comparisons on Clicks}
\label{sec:PRS}
As the observation of documents are position-based and independent \cite{Craswell2008clickpositionbias,wang2018pbe}, treatments on clicked and non-clicked documents will not interfere with each other. Therefore, we can integrate the inverse propensity weighting on the clciked documents and propensity weighting on the non-clicked ones to reweigh the pairwise losses incurred when comparing them:
\begin{align}
\label{eq:PRS}
    &l_{PRS}(\pi_q|q,\tilde\pi_q,o_q) \\\nonumber
    =& \mathop{\sum_{x_i:c_q(x_i)=1}}\sum_{x_j:c_q(x_j)=0}\delta(x_i,x_j|\pi_q)\cdot\frac{P(o_q(x_j)=1|\tilde\pi_q)}{P(o_q(x_i)=1|\tilde\pi_q)},
\end{align}
which leads to a weight on each pairwise loss term defined by the propensity ratio between the non-clicked and clicked documents in it. We name this new weighting scheme Propensity Ratio Scoring, or PRS in short. In expectation, the PRS estimator largely removes the relevant-relevant comparisons, and focus on comparing each relevant document to irrelevant ones for ranker optimization:
\begin{align}
\label{eq:PRS-expectation}
    &E_{o_q}\big[l_{PRS}(\pi_q|q,\tilde\pi_q,o_q)\big] \\\nonumber
    =& E_{o_q}\Bigg[\mathop{\sum_{x_i:o_q(x_i)=1}}_{\wedge\; r_q(x_i)=1}\frac{\sum_{x_j:c_q(x_j)=0}\delta(x_i,x_j|\pi_q)\cdot P(o_q(x_j)=1|\tilde\pi_q)}{P(o_q(x_i)=1|\tilde\pi_q)}\Bigg] \\\nonumber
    =& \sum_{x_i:r_q(x_i)=1}E_{o_q}\bigg[\sum_{x_j:r_q(x_j)=0 \,, o_q(x_j)=1}\delta(x_i,x_j|\pi_q)\bigg] \\\nonumber
    &\;+ \sum_{x_i:r_q(x_i)=1}\; \sum_{x_j:r_q(x_j)=1\,, o_q(x_j)=0}\delta(x_i,x_j|\pi_q)\cdot P(o_q(x_j)=1|\tilde\pi_q)\,,
\end{align}
where the first step is the same as Eq~\eqref{eq:IPS-expectation}; and the second step is derived from Eq~\eqref{eq:PNS-expectation} by substituting $\Omega(x_j|\pi_q)$ with $\delta(x_i,x_j|\pi_q)$. As discussed in Section \ref{sec:PNS}, the second term that consists relevant-relevant comparisons is small and can be safely ignored in practice. In this way, the total loss is merely obtained from valid relevant-irrelevant pairs that contribute to ranker optimization. Notice that besides removing relevant documents from non-clicked documents, Eq \eqref{eq:PRS-expectation} also
excludes the irrelevant and unobserved documents, i.e., the I\&U part in Figure \ref{fig:clickdata}. But as the ranking metrics are defined on relevant documents, we only need to correct the errors introduced by including relevant documents as negative examples, and keep the total loss unbiased to all relevant documents. As demonstrated in the expectation, PRS can properly promote all the relevant documents against irrelevant ones.

Besides the correction effect, we next show that by imposing a more careful use of the non-clicked documents, the variability of the PRS estimator can be reduced comparing to the IPS estimator, which is very important for LTR with real-world click data \cite{adith2015selfnormalized}. Let us first rewrite the PRS estimator in Eq~\eqref{eq:PRS} as follows:
\begin{align*}
\label{eq:PRS-rewrite}
    &l_{PRS}(\pi_q|q,\tilde\pi_q,o_q) \\\nonumber
    =& \mathop{\sum_{x_i:o_q(x_i)=1}}_{\wedge r_q(x_i)=1 }\frac{1}{P(o_q(x_i)=1|\tilde\pi_q)}\cdot E_{o_q}\big[\mathop{\sum_{x_j:r_q(x_j)=0}}_{\wedge o_q(x_j)=1}\delta(x_i,x_j|\pi_q)\big]
\end{align*}
\begin{proposition}
    Let $N$ be the number of relevant documents retrieved for query $q$ and $P(o_q(x_i)=1|\tilde\pi_q)$ be the probability of independent Bernoulli events of observing each relevant document. According to Hoeffding's inequality, for any given new ranking $\pi_q$, with probability of at least $1-\xi$, we have:
    \begin{equation*}
        \Big|l_{PRS}(\pi_q|q,\tilde\pi_q,o_q)-E_{o_q}\big[l_{PRS}(\pi_q|q,\tilde\pi_q,o_q)\big]\Big| \leq \frac{1}{N}\sqrt{\frac{log\frac{2}{\xi}}{2}\sum_{i=1}^{N}\rho_i^2}
    \end{equation*}
    where $\rho_i=\frac{1}{P(o_q(x_i)=1|\tilde\pi_q)}\cdot E_{o_q}\big[\sum_{x_j:r_q(x_j)=0 \wedge o_q(x_j)=1}\delta(x_i,x_j|\pi_q)\big]$ when $0<P(o_q(x_i)=1|\tilde\pi_q)<1$; otherwise, $\rho_i=0$.
\end{proposition}
The complete proof will be elaborated with details in a longer version of this paper. 
The above tail bound of the PRS estimator depicts its variability. Intuitively, this tail bound provides the range that the estimator can vary with a high probability; and a smaller range means a lower variability. Similarly, we can get the tail bound of the IPS estimator in Eq~\eqref{eq:click-unclick} as: 
\begin{equation*}
        \Big|l_{IPS}(\pi_q|q,\tilde\pi_q,o_q)-E_{o_q}\big[l_{IPS}(\pi_q|q,\tilde\pi_q,o_q)\big]\Big| \leq \frac{1}{N}\sqrt{\frac{log\frac{2}{\xi}}{2}\sum_{i=1}^{N}\tau_i^2}
    \end{equation*}
where $\tau_i=\frac{1}{P(o_q(x_i)=1|\tilde\pi_q)}\cdot \sum_{x_j:c_q(x_j)=0}\delta(x_i,x_j|\pi_q)$ if $0<P(o_q(x_i)=1|\tilde\pi_q)<1$; and $\tau_i=0$, otherwise. As the propensity of each non-clicked document is smaller than 1, $0\leq\rho_i\leq\tau_i$ always holds. Thus, the PRS estimator enjoys a reduced variability than IPS, which is vital for the convergency of ranker estimation in practice.

In addition to providing an unbiased estimate of pairwise comparisons on click data, another obvious advantage of PRS is its general applicability: it does not require any additional statistics or procedures than those already used in IPS (e.g., we can use any existing methods for propensity estimation \cite{joachims2017ips,Craswell2008clickpositionbias,wang2018pbe,Agarwal2019EPB,attribute}). As a result, it can be seamlessly applied to all existing unbiased LTR settings or other scenarios (e.g., recommendation \cite{Schnabel2016RecTreat}) where IPS is used, and guaranteed for better performance.

So far, we have assumed a noise-free setting, i.e., a document is clicked if and only if it is observed and relevant: $c_q(x_i)=1\Leftrightarrow [o_q(x_i)=1\wedge r_q(x_i)=1]$. 
However, in reality this may not hold: a user can possibly misjudge and miss a relevant document, or mistakenly click on an irrelevant document. Fortunately, PRS is order-preserving under the same noise assumption made by IPS \cite{joachims2017ips}. 
Due to space limit, we decide not to include the proof, which can be obtained similarly as in \cite{joachims2017ips} but based on Eq \eqref{eq:PRS-expectation}. We will present the influence of noisy clicks empirically in Section \ref{sec:exp_synthetic}. 

\section{Evaluation}
\label{sec:exp}
In this section, we conduct comprehensive empirical evaluations of PRS for unbiased LTR. We apply PRS to different ranking algorithms to show its wide applicability. We first synthesize the clicks following the conventional procedure \cite{joachims2017ips,ai2018DLA,hu2019unbiasedlambda} on three benchmark LTR datasets to study the behaviors of PRS from different perspectives. To confirm the effectiveness of PRS in an industrial setting, we also perform experiments on the large-scale GMail search data. 

\subsection{PRS on Different Ranking Models}
As shown in Eq~\eqref{eq:PRS}, the proposed PRS estimator generally applies to any pairwise LTR algorithms. To test the performance of PRS on different ranking models, we include two popularly used yet significantly different pairwise LTR algorithms for experiments. One is pairwise logistic regression \cite{rendle2009bpr,Arias2008logistic}, and the other one is LambdaMART \cite{burges2010from}, the state-of-the-art pairwise LTR algorithm.

\noindent$\bullet$ \textbf{Pairwise Logistic Regression.} It uses a logistic function to measure the likelihood of $x_i$ being more relevant than $x_j$ under query $q$, based on their predicted relevance.
By taking the logarithm of the logistic function, the pairwise logistic loss for optimization is:
\begin{equation*}
        \delta(x_i,x_j|\pi_q) = \log\big(1+e^{-\big(\tilde r_q(x_i)-\tilde r_q(x_j)\big)}\big)
\end{equation*}
We use a linear scoring function $\tilde r_q(x_i)=\omega^\top\phi(x_i,q)$, where $\phi(x_i,q)$ is a feature vector that describes the matching between $x_i$ and $q$. We can directly substitute the above loss in Eq~\eqref{eq:PRS} to get the PRS estimator for pairwise logistic regression and add an $l_2$ regularization to control overfitting.

\noindent$\bullet$ \textbf{LambdaMART.} It combines MART \cite{Friedman2000mart} and the lambda functions from LambdaRank \cite{Burges2006lambdarank}. Its optimization is directly performed with respect to the lambda functions. 
To apply PRS with click data in LambdaMART, we first denote a set of document pairs $I_q=\{(x_i,x_j)|c_q(x_i)=1\wedge c_q(x_j)=0\}$ in each query $q$. Then we modify the lambda functions as:
\begin{align*}
    &\tilde\lambda_i = \sum_{j:(x_i, x_j)\in I_q}\tilde\lambda_{ij}-\sum_{j:(x_j,x_i)\in I_q}\tilde\lambda_{ji}\\
    \text{where}\,\;&\tilde\lambda_{ij} = \frac{-\sigma|\Delta Z_{ij}|}{1+e^{\sigma(\tilde r_q(x_i)-\tilde r_q(x_j))}}\cdot\frac{P(o_q(x_j)=1|\tilde\pi_q)}{P(o_q(x_i)=1|\tilde\pi_q)}
\end{align*}
$\Delta Z_{ij}$ is the change of the ranking metric of interest (e.g., NDCG) if documents $x_i$ and $x_j$ are swapped in the ranked list $\pi_q$; $\sigma$ is a global weighting coefficient. $\tilde\lambda_{ij}$ without the PRS weight can be viewed as the pairwise loss $\delta(x_i, x_j|\pi_q)$ in Eq~\eqref{eq:PRS}. We applied the modified lambda functions in the standard LambdaMART implementation, which we name as PRS-LambdaMART. 

We primarily compare the PRS estimator against the IPS estimator, and also a Naive estimator. The Naive estimator simply treats clicks as relevance and non-clicks as irrelevance. The two baseline estimators can be viewed as special cases of our PRS estimator: IPS always sets the observation propensity of a non-clicked document to $1$; and Naive simply sets the PRS weight to $1$ in all pairs. All three estimators are applied to the two base ranking algorithms to learn new rankers from click data. We also include the Full-Info ranker trained with fully-labeled ground-truth data as the skyline. 
\subsubsection{Reasoning of Weight Clipping}
Similar to the clipping trick used in IPS \cite{joachims2017ips}, proper clipping on the PRS weights is important to safeguard its stable performance in real applications. The main reason of weight clipping is for variance reduction, and we found the variance highly depends on the quality of the ranker that presents the logged ranking $\tilde\pi_q$ (referred to as production ranker in \cite{joachims2017ips}). 
Specifically, 
when a low-quality production ranker is used when collecting clicks, the chance that some irrelevant documents ranked at the top and relevant documents ranked at the bottom will be high. Hence, there will be pairs in which a bottom ranked relevant document is clicked and a top ranked irrelevant document is not. As the PRS weight is the ratio between the propensities of the non-clicked and clicked documents, such pairs will have very large PRS weights and thus dominate the total loss. 
The resulting ranker will then be forced to correct these edge cases, e.g., go against or even reverse the production ranker, but totally miss other documents. In practice, to avoid such effect, we clip the PRS weights in Eq~\eqref{eq:PRS} with a constant $\gamma$ (only clipping the small propensities of lower-ranked clicked documents as in IPS \cite{joachims2017ips} yields similar effect):
\begin{align}
\label{eq:PRS-clip}
    &l_{PRS}(\pi_q|q,\tilde\pi_q,o_q) \\\nonumber
    =& \mathop{\sum_{x_i:c_q(x_i)=1}}\sum_{x_j:c_q(x_j)=0}\delta(x_i,x_j|\pi_q)\cdot \min\bigg\{\gamma,\frac{P(o_q(x_j)=1|\tilde\pi_q)}{P(o_q(x_i)=1|\tilde\pi_q)}\bigg\}
\end{align}
Empirically, we set $\gamma=1$ which is found to be effective in our experiments. However, when the ranking model used to train a new ranker has sufficient capacity (e.g., a non-linear model) to fit all pairs including both the extreme and regular ones, the influence of those large weights becomes less a concern and we can relax the clipping. For example, LambdaMART is less sensitive to the extreme pairs than Logistic Regression in our observations. 
For the IPS estimator, the same analysis applies and we follow the clippings suggested in \cite{joachims2017ips} to make a fair comparison. 
\begin{figure*}[ht]
    \centering
        \includegraphics[width=0.31\linewidth]{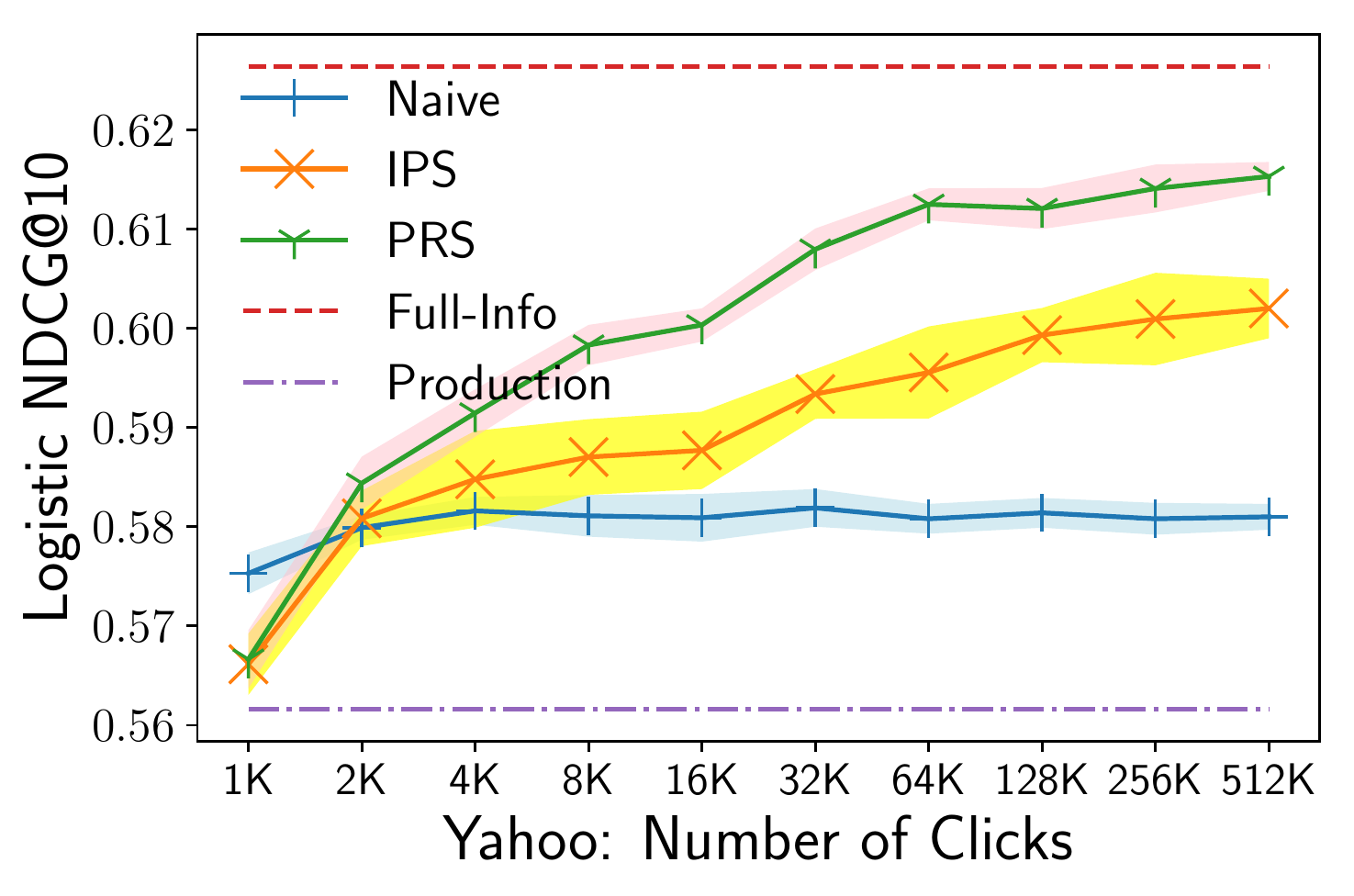}
        \includegraphics[width=0.31\linewidth]{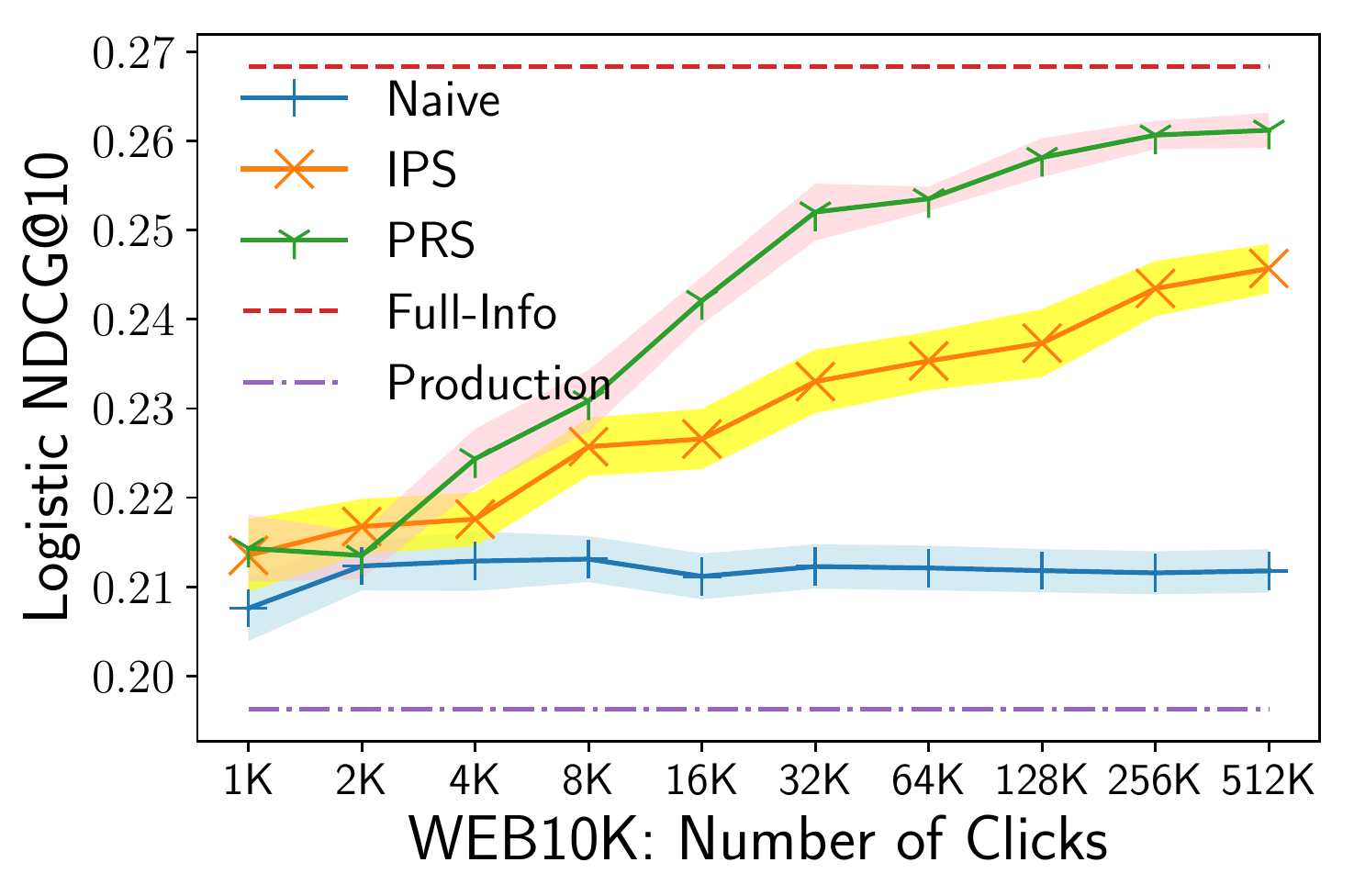}
        \includegraphics[width=0.31\linewidth]{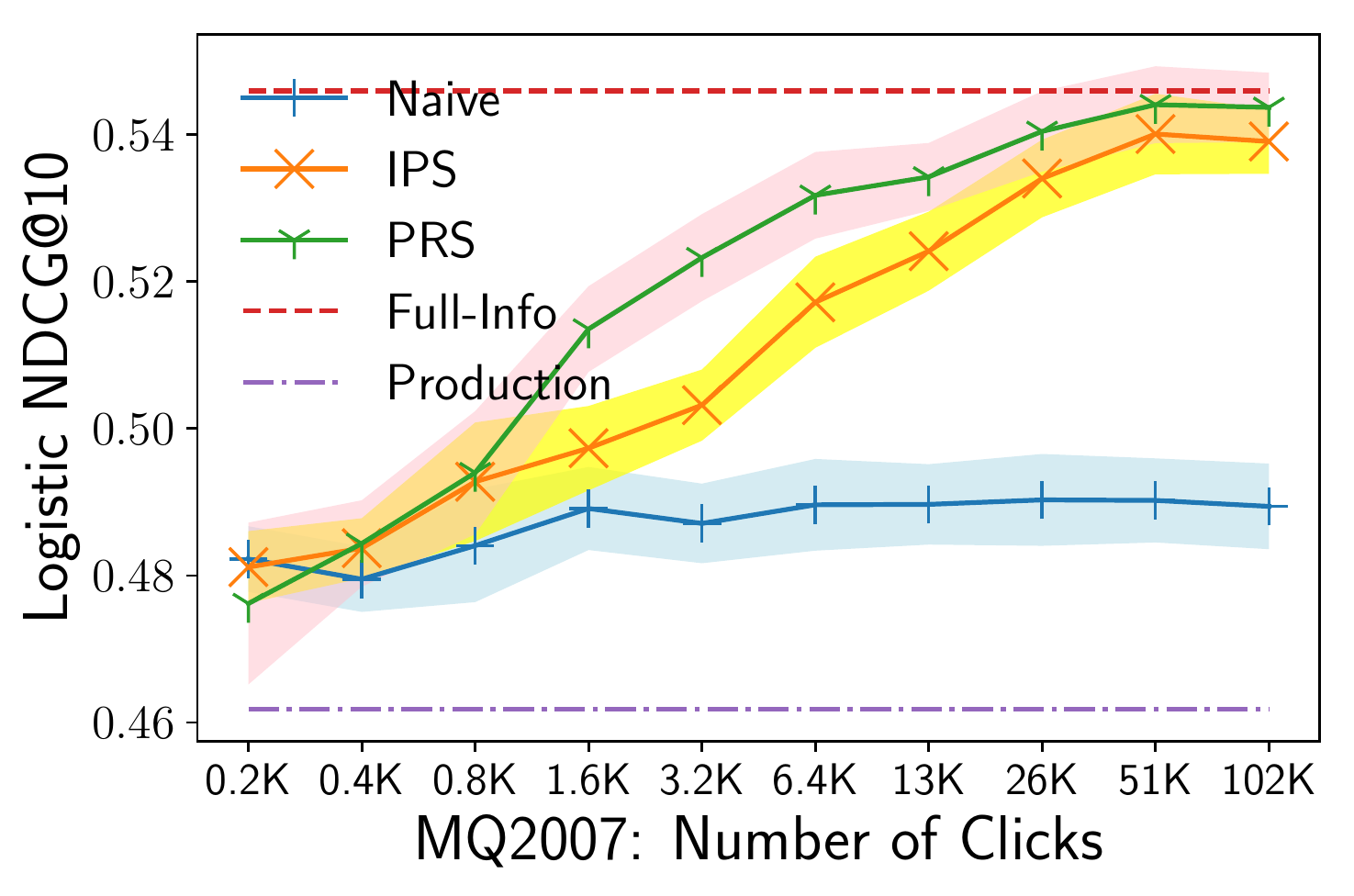}
    \centering
        \includegraphics[width=0.31\linewidth]{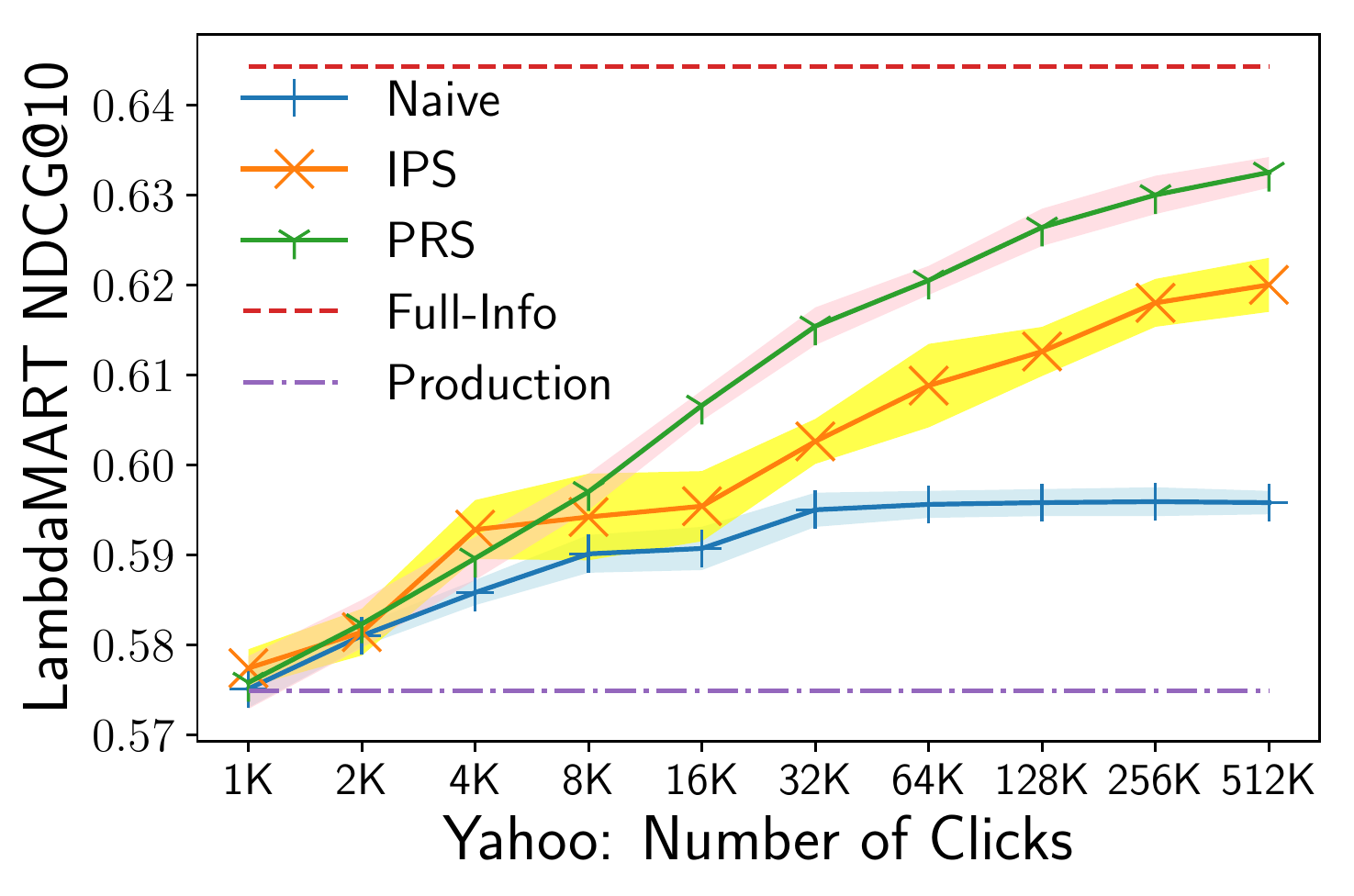}
        \includegraphics[width=0.31\linewidth]{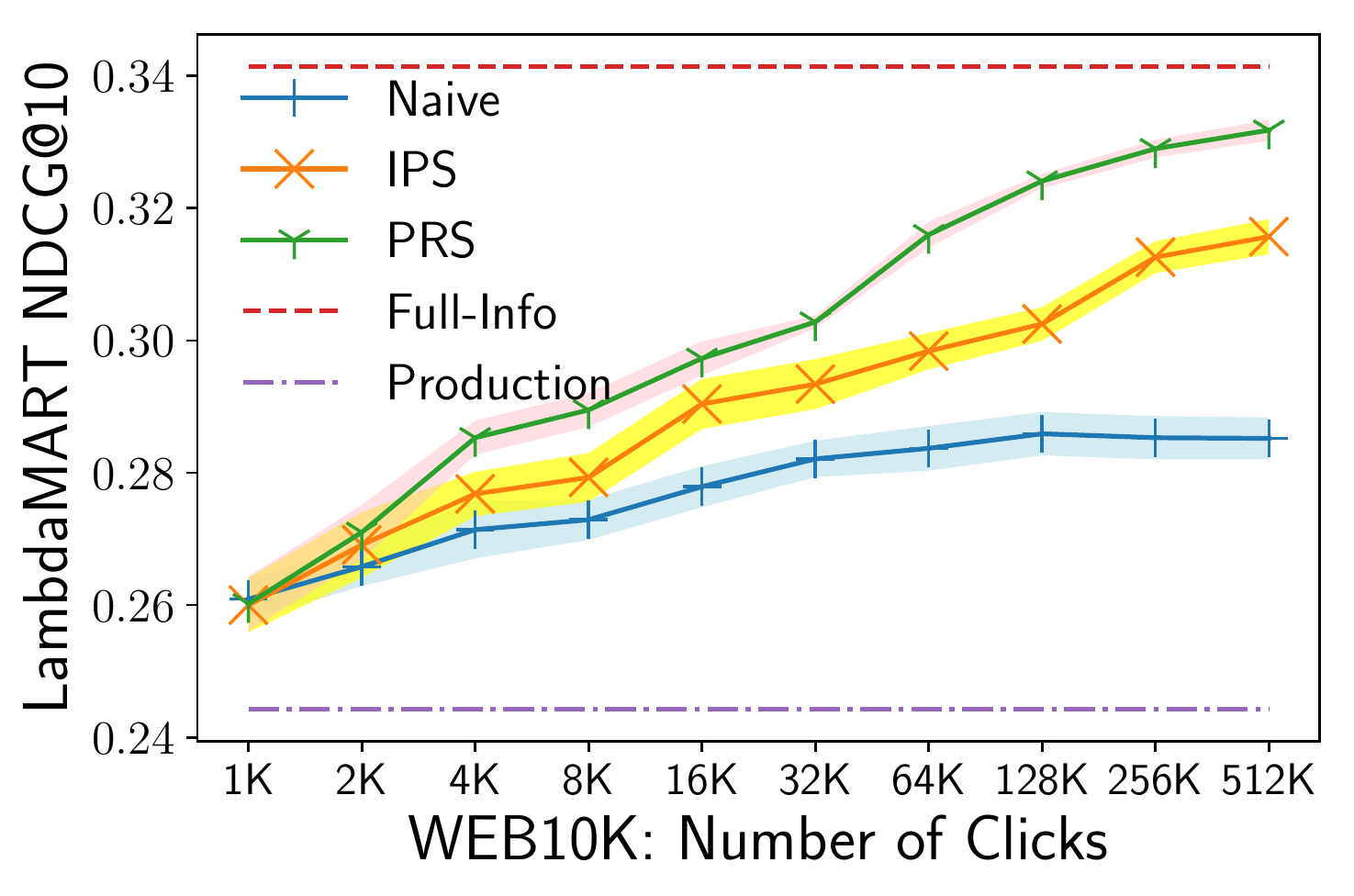}
        \includegraphics[width=0.31\linewidth]{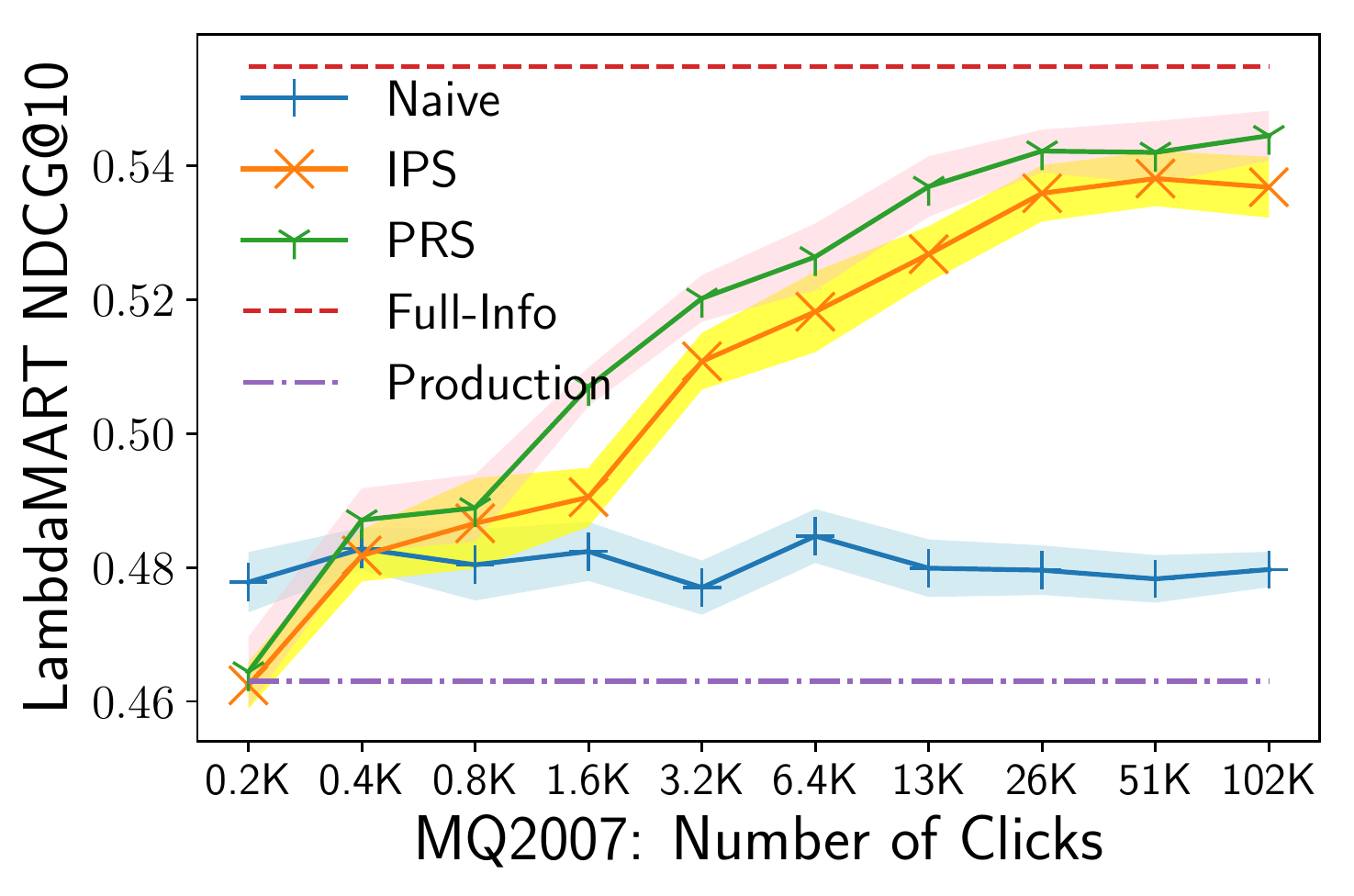}
    \vspace{-0.5em}    
    \caption{The test set ranking performance of different rankers. The results are from the clicks on two base ranking algorithms over three datasets as indicated in the figures. The shadow areas denote standard deviations at each result point. ($\eta=1,\mu=0.1$)}
    \label{fig:click_scale}
\end{figure*}
\subsection{Synthesize Clicks on LTR Benchmarks}
\label{sec:synthesize}
We adopt three benchmark LTR datasets, including Yahoo Learning to Rank Challenge (set1), MSLR-WEB10K and MQ2007.\\
$\bullet$ \textbf{Yahoo.}
One of the largest and most popularly used benchmark dataset for LTR. It contains around $30K$ queries with $710K$ documents. Each query-document pair is depicted with a 700-dimension feature vector (519 valid features) and a five-grade relevance label. Following \cite{joachims2017ips}, we binarize the relevance by assigning $r_q(x)=1$ to documents with relevance label $3$ or $4$, and $r_q(x)=0$ for others. \\
$\bullet$ \textbf{WEB10K.} 
It contains $10K$ queries and a 136-dimension feature vector for each query-document pair. We binarize the relevance labels in the same way as in the Yahoo dataset. \\
$\bullet$ \textbf{MQ2007.}
It contains about $1,700$ queries and a 46-dimension feature vector for each query-document pair, with relevance label in $\{0,1,2\}$. We assign $r_q(x)=1$ to documents with relevance label $1$ or $2$, and $r_q(x)=0$ to documents with relevance label $0$. 

For all three datasets, we keep the partition of the train, validation, test set from the corpus and report the performance on the binarized fully labeled test sets with five-fold cross-validation. We follow the procedure in \cite{joachims2017ips} to derive click data from each fully labeled dataset. To generate clicks, we first train an initial ranker with 1 percent of the queries in the training set, which is referred to as the production ranker $\pi_0$. Then we randomly select a query $q$ from the rest of the training set, for which we use $\pi_0$ to compute the ranking $\tilde\pi_q$ for this query. With the ranked list, we can generate clicks according to a position-based click model,
\begin{equation*}
    P(c_q(x_i)=1|\tilde\pi_q) = P(e_q(x_i, k)=1|\tilde\pi_q)\cdot P\big(r_q(x_i)=1|e_q(x_i, k)=1\big),
\end{equation*}
where $e_q(x_i, k)$ denotes whether the document at position $k$ is examined, and examination equals the observation of the relevance label: $e_q(x_i) = o_q(x_i)$. Thus the observation propensity is equivalent to the position-based examination probability $P(e_q(x_i,k)=1|\tilde\pi_q)$, 
which is defined as follows,
\begin{equation*}
    P(o_q(x_i)=1|\tilde\pi_q) = p_{rank(x_i|\tilde\pi_q)}=\bigg(\frac{1}{rank(x_i|\tilde\pi_q)}\bigg)^\eta
\end{equation*}
where $\eta$ represents the severity of position bias. The propensity $P(o_q(x_i)=1|\tilde\pi_q)$ of each document (regardless of clicked or non-clicked) is recorded based on their positions in the presented ranking. 
We also add click noise as in \cite{joachims2017ips}. Denote $\mu\in[0,0.5)$ as the noise level, we have $P\big(c_q(x_i)=1\big|r_q(x_i)=1,o_q(x_i)=1\big)=1-\mu$ and $P\big(c_q(x_i)=1\big|r_q(x_i)=0,o_q(x_i)=1\big)=\mu$. When not mentioned otherwise, we use $\eta=1$ and $\mu=0.1$ as the default setting. In the following sections, we investigate the influence of each component. 

We use NDCG@10 as the main performance metric. We also computed other metrics such as MAP and ARP. But as the performance on these metrics was consistent with each other and the space limit, we only report NDCG@10 to include more experiments. We tune the hyper-parameters via cross-validation. Each result is averaged over five runs and the standard deviation is displayed as the shadow areas in the figures.

\subsection{Evaluations on Synthetic Data}
\label{sec:exp_synthetic}
\subsubsection{Performance with the Scale of Click Data}
We first study how the ranking performance scales with the number of clicks. The results are reported in Figure \ref{fig:click_scale}, and the production ranker $\pi_0$ is used as a baseline. The x-axis denotes the number of clicks, and the y-axis reports the NDCG@10 on the fully labeled test set. The figures show that PRS consistently and significantly outperforms IPS with an increasing size of the click data, in both ranking algorithms across three datasets. When there are only a few clicks, the ranking model cannot be sufficiently estimated. Hence the size of training data is the major bottleneck rather than the bias in it.
But with an increasing number of clicks, IPS introduces more relevant documents from the non-clicked part into the comparisons, which distorts the optimization of rankers. Besides, the variance of PRS is reduced considerably due to a finer-grained use of the non-clicks. The Naive estimator does not consider the bias at all and thus cannot make effective use of click data. This experiment demonstrates clear advantages of PRS in our default setting. Next we will focus on specific perspectives of the unbiased LTR settings. Due to the space limit, we only report the results on Yahoo dataset, but the conclusions are consistent with the other two datasets. 
\begin{figure}
    \centering
    \includegraphics[width=0.495\linewidth]{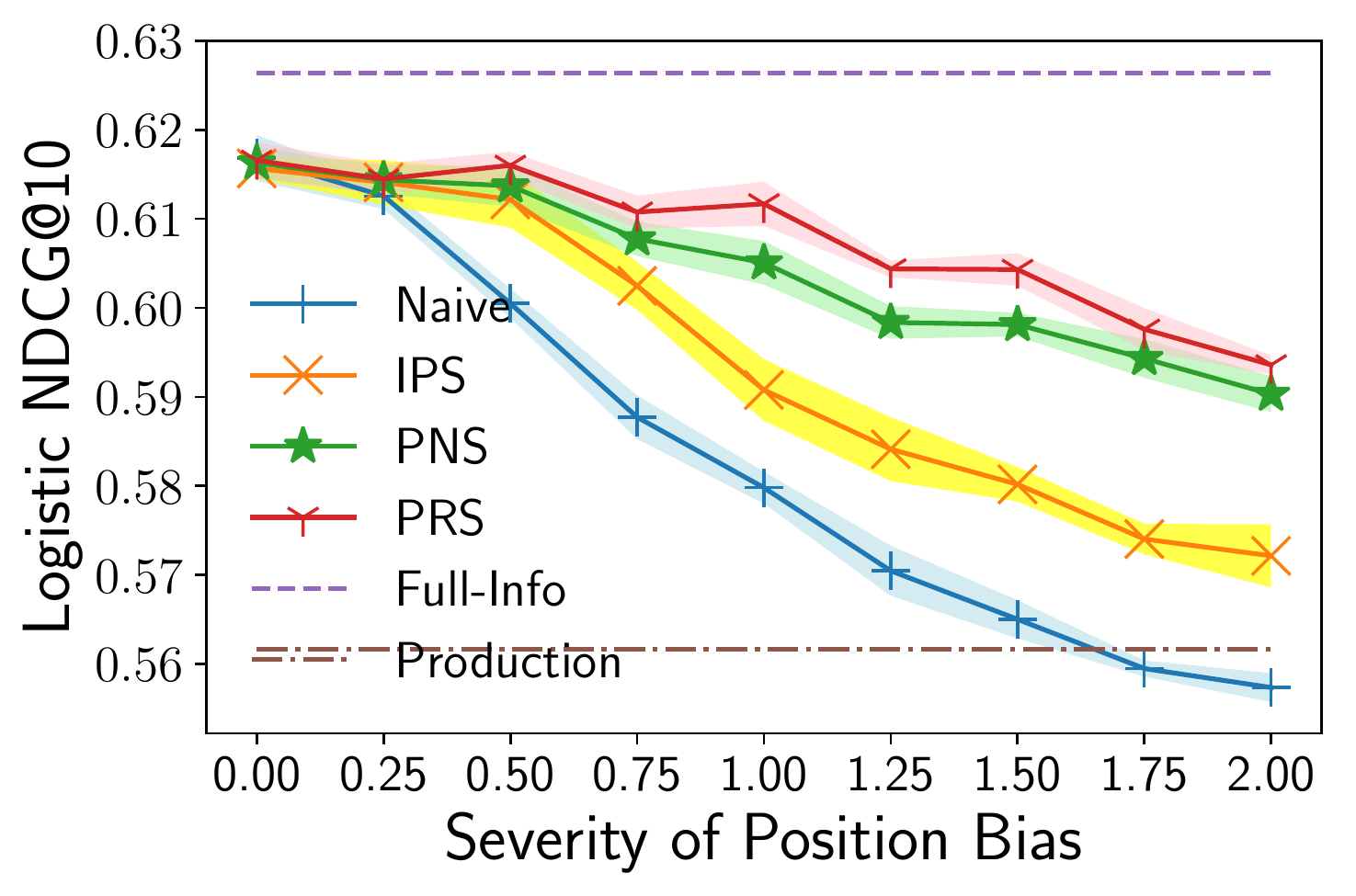}
    \includegraphics[width=0.495\linewidth]{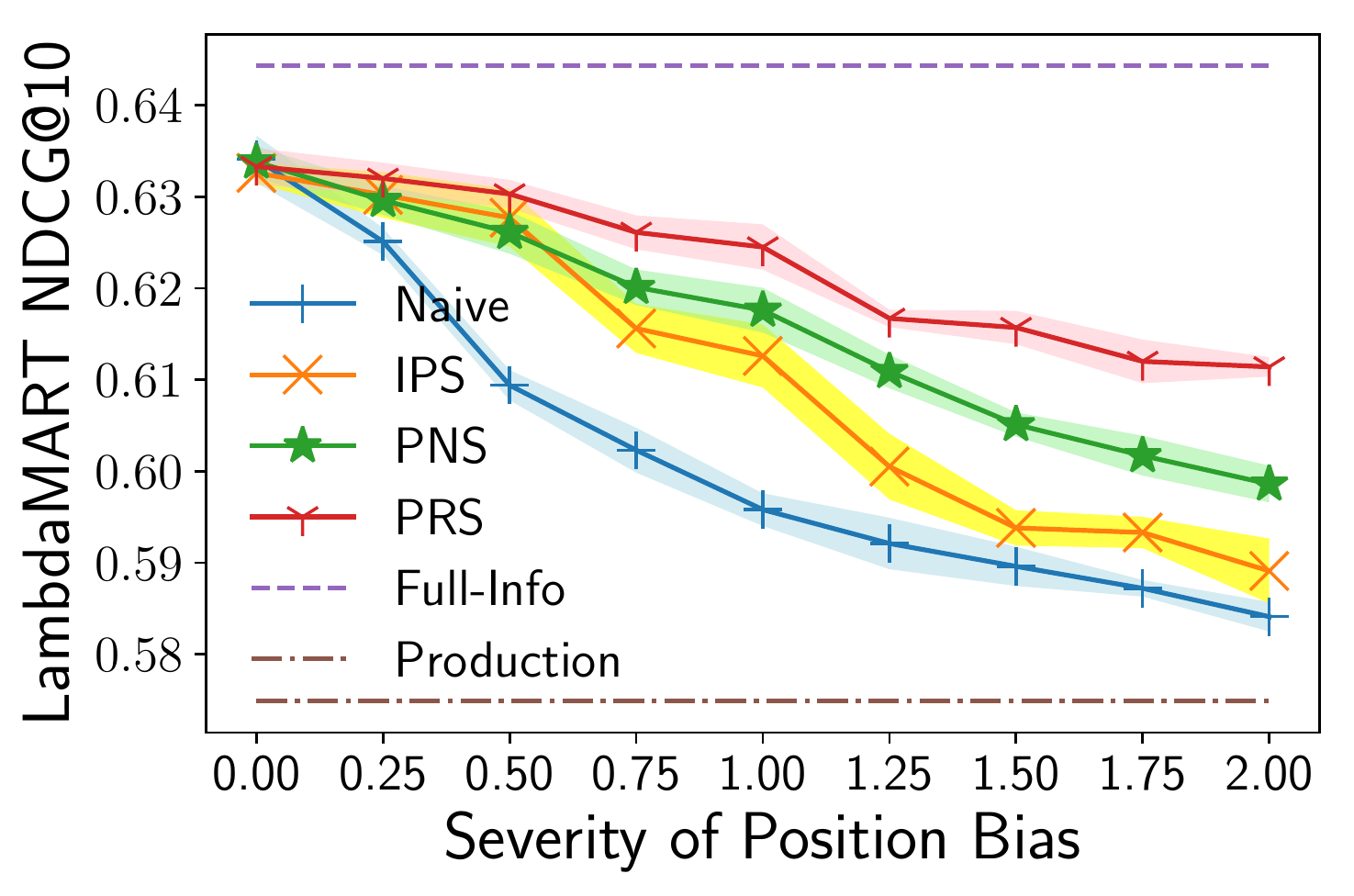}
    \vspace{-1.5em}
    \caption{The performance of different rankers under different degree of position bias controlled by $\eta$. ($n=128K,\mu=0.1$)}
    \label{fig:bias-severity}
    \vspace{-4mm}
\end{figure}

\subsubsection{Tolerance to the Severity of Position Bias}
We investigate the performance of the PRS estimator under different degrees of position bias in clicks. We vary $\eta$ from $0$ to $2$ when generating clicks. In order to better understand the effect of position bias, we also include the weighting solely on non-clicked documents (PNS) as introduced in Section \ref{sec:PNS}. PNS focuses on identifying truly irrelevant documents in unclicked documents, but does not correct the bias on clicks. 
Figure \ref{fig:bias-severity} demonstrates the influence of the bias on different estimators for LTR. PRS achieves the best performance as it properly handles both clicks and non-clicks. The IPS estimator only works when there is a low level of position bias, while its advantage over the Naive estimator diminishes with increased bias. This is consistent with the previously reported findings in \cite{Jagerman2019modelorintervene}. It is worth noting that the PNS estimator is more robust to stronger position bias than IPS and Naive. This observation suggests that the consequence of including relevant documents 
as negative examples can be even more severe than the bias issue in clicked documents.
\subsubsection{Robustness to Click Noise}
\begin{figure}
    \centering
    \includegraphics[width=0.49\linewidth]{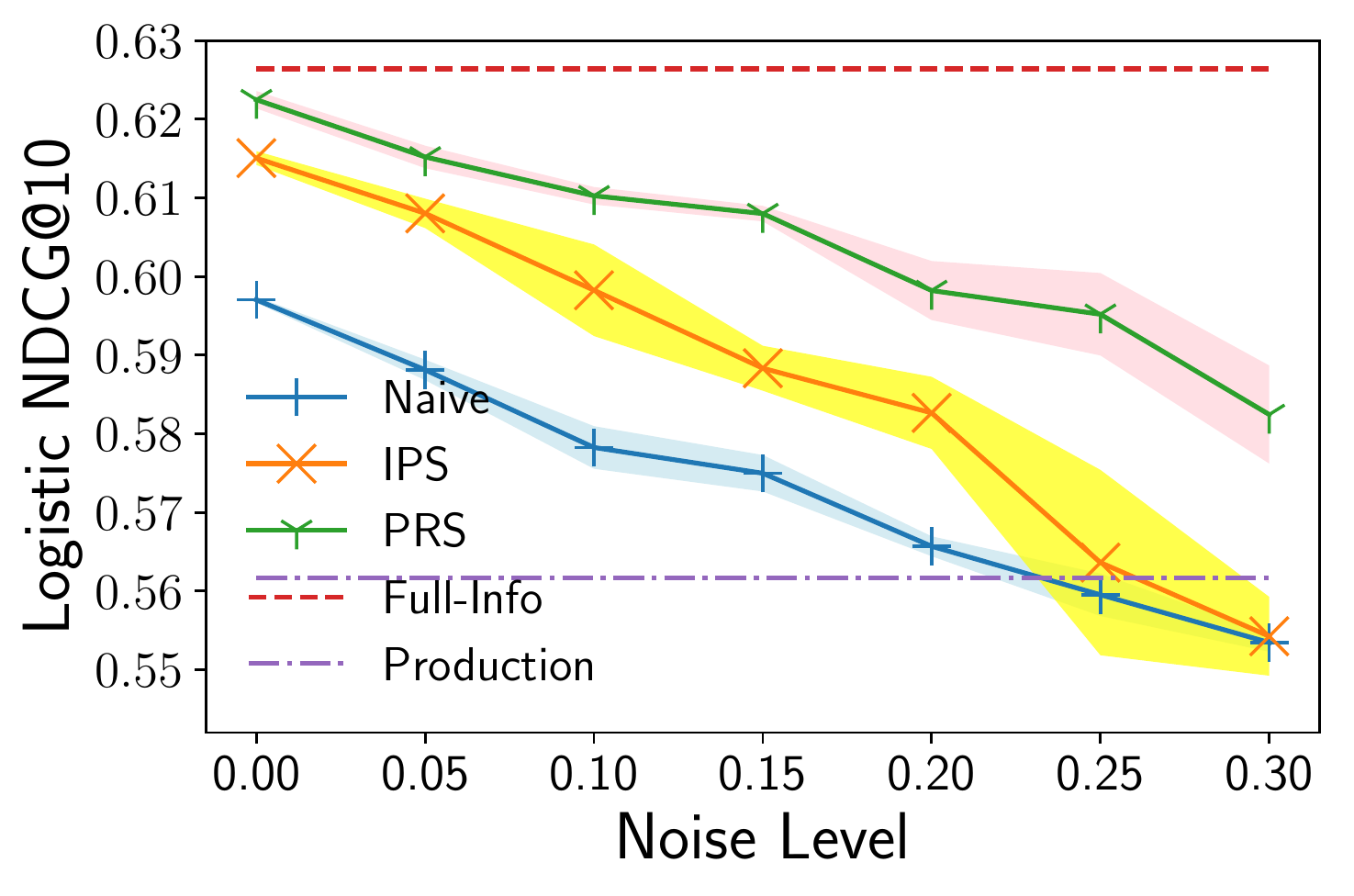}
    \includegraphics[width=0.49\linewidth]{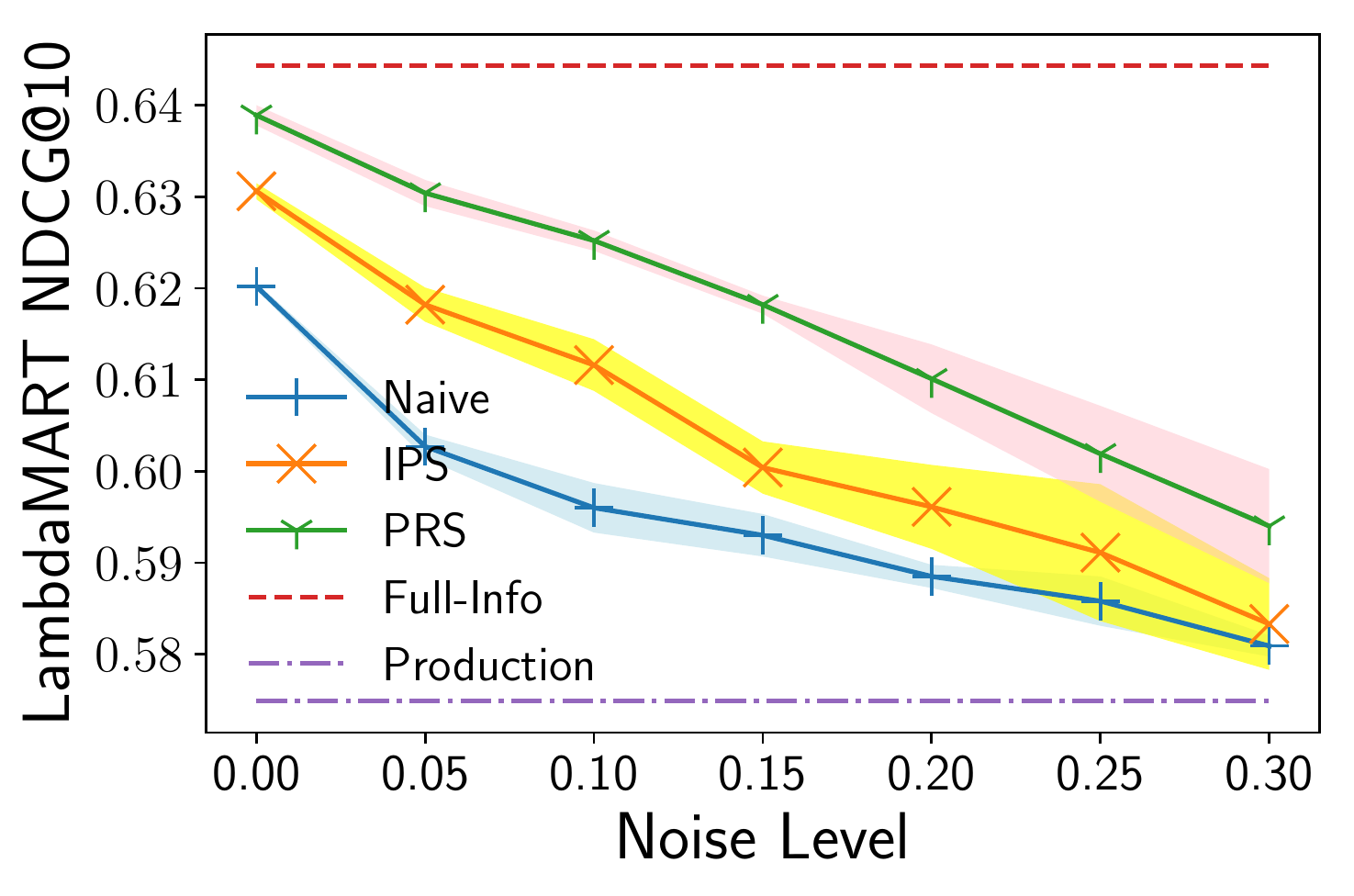}
    \vspace{-1em}
    \caption{The performance of different rankers with an increasing click noise level ($n=128K, \eta=1$).}
    \label{fig:noise-level}
\end{figure}
We now evaluate the robustness of different estimators to click noise, by varying the noise level $\mu$ in Figure \ref{fig:noise-level}. The results show that PRS is more resistant to click noise. This is again due to its treatments on both clicked and non-clicked documents. For example, when an irrelevant document is clicked, its large IPS weight will be canceled by the propensity weight on the non-clicked document introduced by PRS. And therefore, this erroneous pair generates less impact on ranker estimation. In comparison, when the noise level is high, the performance of IPS can drop to the Naive estimator's performance, because of its unreasonably high weight on the erroneous pairs. 

\subsubsection{Robustness to Misspecified Propensities}
\begin{figure}
    \centering
    \includegraphics[width=0.49\linewidth]{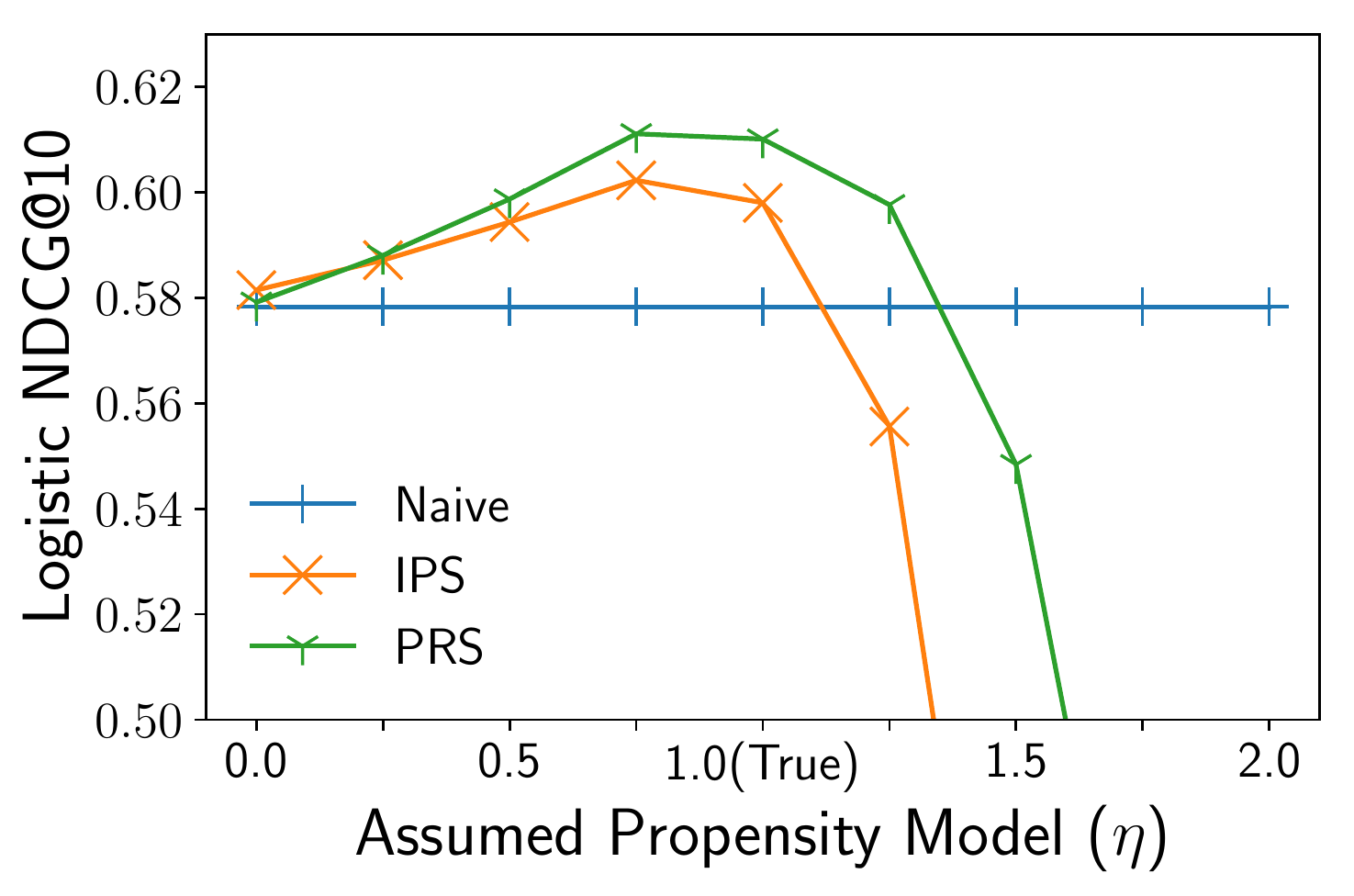}
    \includegraphics[width=0.49\linewidth]{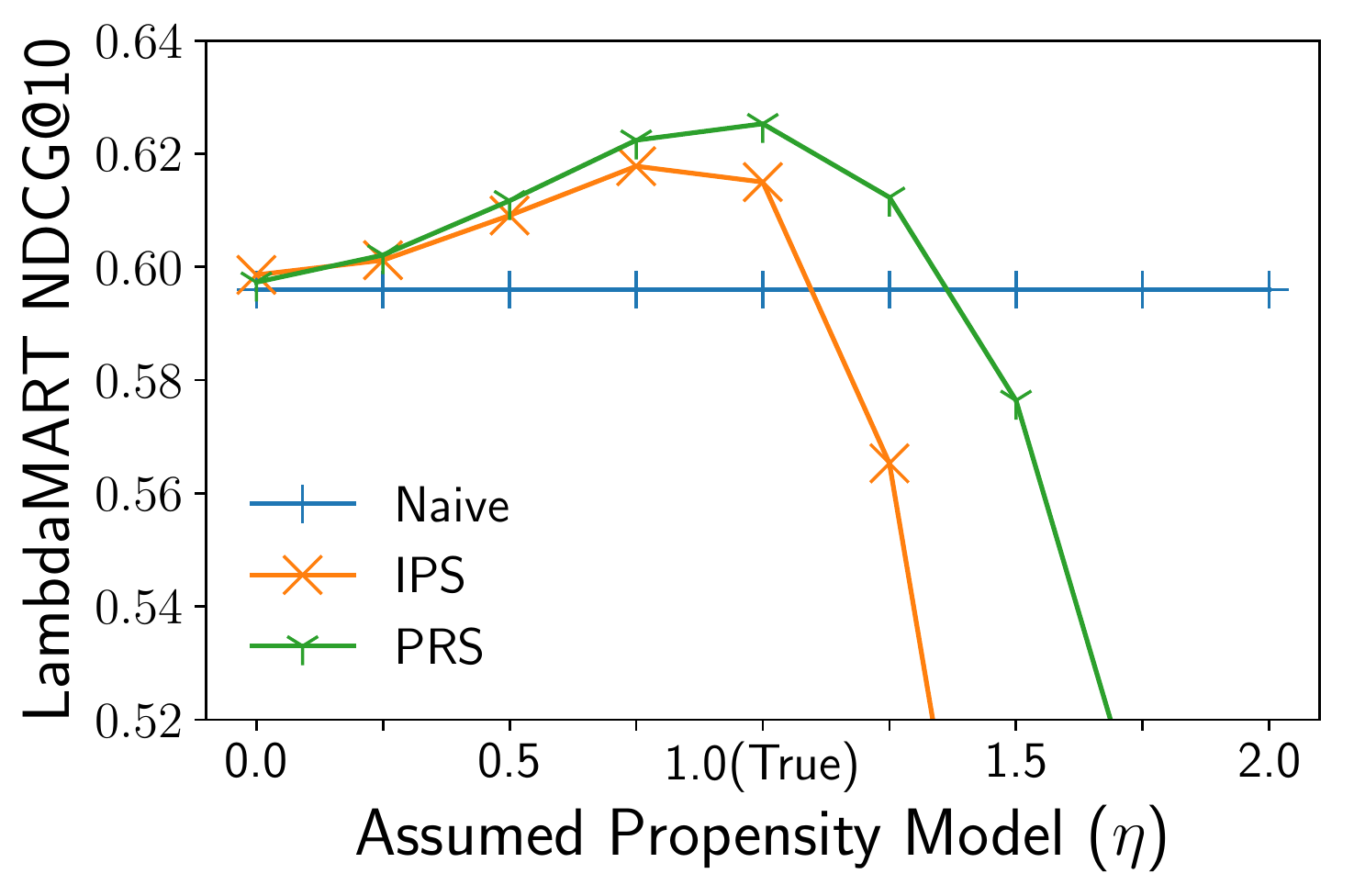}
    \vspace{-1em}
    \caption{The performance of different rankers with misspecified propensities ($n=128K,\; \text{true}\;\eta=1,\;\mu=0.1$).}
    \label{fig:mis-propensity}
    \vspace{-1em}
\end{figure}
We have assumed the access to accurate propensities. However, this is not always the case in practice and the propensities may need to be estimated based on a specified propensity model. In this experiment, we evaluate how robust different estimators are to misspecified propensities. We use $\eta=1$ to generate clicks, but with different $\eta$ in training. The results are shown in Figure \ref{fig:mis-propensity}, where x-axis is the $\eta$ for the assumed (estimated) propensities in click data. Both IPS and PRS are less sensitive to the overestimated propensities (when $\eta<1$). But PRS is much more robust than IPS when the propensities are underestimated (when $\eta>1$). This is because in PRS, as long as the propensity \emph{ratios} are close to those from the true propensities, the ranker estimation quality could be largely maintained. This analysis also illustrates the practical advantage of PRS when accurate propensities are difficult to obtain. 

\subsubsection{Comparisons with Variants of IPS}
We use PRS-LambdaMART as our solution for unbiased LTR, and compare it with recent variants of IPS-based methods. The first one we include is Propensity SVM-Rank \cite{joachims2017ips}, which applies IPS to SVM-Rank. The second baseline is the Dual Learning Algorithm (DLA) \cite{ai2018DLA}. It 
uses a DNN model to jointly estimate propensities and an unbiased ranker from click data as a dual problem. The last one is Unbiased LambdaMART \cite{hu2019unbiasedlambda}, which applies IPS weights to both clicked and non-clicked documents and jointly estimates the propensities and the ranker in a pairwise manner. For a fair comparison, we estimate the propensities with randomization from the generated clicks as used in \cite{ai2018DLA,hu2019unbiasedlambda} for our solution. 
The comparisons of the ranking performance are shown in Table \ref{tab:ips-variants} under paired t-test with $p$-value$<$0.05. First, pairwise comparisons between clicked and non-clicked documents used in Unbiased LambdaMART and PRS-LambdaMART are more effective than comparing each clicked documents to all others as in the other two baselines. The Unbiased LambdaMART algorithm can be understood as a special case of our PRS scheme. It applies the IPS weights to non-clicked documents as $\frac{1}{t^-}$, by assuming the non-click probability is proportional to the irrelevance probability. As $t^-$ is not bounded, i.e., it can be larger than 1, $\frac{1}{t^-}$ could achieve a similar effect as the propensity weight on non-clicked documents. But as $t^-$ is not bounded and 
largely relies on the regularization for estimation, there is no guarantee it can recover the correct propensity and achieve the desired effect. This leads to its worse performance. On the other hand, this also shows the possibility of generalizing PRS to jointly learning propensities and the unbiased ranker.
\begin{table}[]
\caption{The ranking performance of different unbiased LTR solutions ($n=128K,\eta=1,\mu=0.1$).} 
\vspace{-1em}
\label{tab:ips-variants}
\begin{tabular}{@{}cccc@{}}
\toprule
                    & MAP             & NDCG@5          & NDCG@10         \\ \midrule
Propensity SVM-Rank & 0.5720          & 0.5877          & 0.5978          \\
DLA                 & 0.5785          & 0.5974          & 0.6009          \\
Unbiased LambdaMART & 0.5961          & 0.6121          & 0.6173          \\
PRS-LambdaMART      & \textbf{0.6049}$^*$ & \textbf{0.6206}$^*$ & \textbf{0.6264}$^*$ \\ \bottomrule
\end{tabular}
\vspace{-0.75em}
\end{table}

\subsection{Evaluations on GMail Search Data}
We further evaluate PRS on data from one of the world's largest personal search engines, GMail search, to prove PRS' applicability in real-world large-scale industrial settings. Specifically, we collected the click-through data from Gmail search logs between June and July 2020 for experiments, resulting in hundreds of millions of queries with clicks. The ranking order of documents in each query was determined by the actual deployed production model, and other factors such as (the unknown) click noise are without any synthetic interventions. Among all the queries, 80\% are used for training, 10\% for validation and parameter tuning, and the remaining for testing. On average, each GMail query has six candidate documents based on its search interfaces and one of the documents is clicked.

The dataset contains a rich set of features, including query-document matching features such as BM25, situational features (e.g. time of the day), user features (e.g. user’s previous click behavior), and document attributes (e.g. document age), etc. The observation propensities are estimated from randomized online experiments \cite{Wangselectionbias}.
We use weighted mean reciprocal rank (WMRR) as our primary metric, since it has been found to be predictive of online gains \cite{wang2018pbe}. In particular, WMRR is calculated as follows: 
\begin{equation}
\mathrm{WMRR} = \frac{1}{\sum_{i=1}^{N} w_i} \cdot \sum_{i=1}^N w_i \frac{1}{\mathrm{rank}_i},
\end{equation}
where $w_i$ denotes the bias correction weight that corresponds to the inverse propensity of the clicked document, $N$ denotes the number of testing queries, $\mathrm{rank}_i$ denotes the rank position of the clicked document for the i-th query. As explained in Section \ref{sec:unbiaLTR}, WMRR is an unbiased estimate of MRR metric. Hence, the higher WMRR is, the better a model performs.

We conduct experiments with linear rankers, LambdaMART, and deep neural network (DNN) rankers to demonstrate PRS' robust performance on real-world datasets with different families of ranking models. We include the following specific models for comparison: a baseline logistic regression model using Naive estimator, an IPS-trained logistic regression, a PRS-trained logistic regression, Unbiased LambdaMART (based on IPS) \cite{hu2019unbiasedlambda}, PRS-LambdaMART, an IPS-trained DNN model, and a PRS-trained DNN model. For all compared models, we use the same set of input features. For DNN models, we use a 3-layer fully connected architecture with 256 hidden dimensions, pairwise logistic loss, and SGD for optimization. All results are reported comparatively to the base logistic regression model, to avoid reporting the absolute performance that is proprietary. As a large portion of testing queries are already clicked at the top (position 0), which are considered as easier queries, we also report the results on other queries separately to better illustrate the performance improvement. The results are shown in Table \ref{tab:gmail-linear}. Note that in large commercial search systems, an approach with an improvement around 0.2\% is considered as substantial \cite{Wangselectionbias, mcn}.

\begin{table}[]
\caption{Ranking performance on GMail data using WMRR.} \small
\vspace{-1em}
\label{tab:gmail-linear}
\begin{tabular}{@{}cccc@{}}
\toprule
                    & Overall             & Clicked at top          & Clicked at others         \\ \midrule
IPS-linear                 & +0.11\%          & -0.19\%          & +0.40\%          \\
PRS-linear                 & +0.39\%          & +0.13\%          & +0.61\%          \\
Unbiased LambdaMART        & +3.92\%          & +5.54\%          & +0.77\%          \\
PRS-LambdaMART             & +4.14\%          & +5.71\%          & +0.99\% \\
IPS-DNN        & +3.91\%          & +5.65\%          & +0.64\%          \\
PRS-DNN             & +4.11\%          & +5.69\%          & +0.96\% \\\bottomrule
\end{tabular}
\vspace{-0.75em}
\end{table}

All relative differences between PRS and baseline in each family of algorithms are statistically significant under paired t-test with $p$-value$<$0.05. We can find that PRS not only performs the best among all solutions, it is also robust under different queries, including those more difficult ones (i.e., clicked on lower ranked positions). IPS tends to over-emphasize difficult queries (e.g., give them larger weights in training), while sacrificing performance on easier ones. LambdaMART and DNN rankers outperform linear rankers by a large margin due to their model capacity in leveraging feature non-linearity. Unbiased LambdaMART shows competitive performance, but is still inferior to PRS-LambdaMART on this large-scale real-world dataset. The results also strongly support PRS for real-world deployments: it only requires a simple modification of the instance weighting schema during training, without any other changes (e.g., logging or optimization).  
\section{Conclusions and Future Work}


In this work, we identify and prove the deficiency of IPS in unbiased LTR tasks. In particular, IPS inevitably includes the relevant-relevant document comparisons when using click data for LTR, which distorts ranker optimization. We instead propose a new weighting scheme named PRS that imposes treatments on both clicks and non-clicks. Comprehensive empirical evaluations on both synthetic and real-world Gmail search data demonstrate the significance of PRS for practical use. 

This paper lays the theoretical basis of effectively utilizing implicit feedback, such as clicks, for LTR. The proposed PRS solution can be easily applied to where IPS is used without any infrastructure change. Although PRS is developed based on pairwise comparisons, it is straightforward to generalize to pointwise LTR methods.
Besides, recent advances in listwise LTR also indicate the importance of comparing each document to its less relevant peers \cite{zhu2020listwise}. It is important to extend the idea of removing relevant documents from non-clicks in PRS for effective listwise LTR.

\begin{acks}
We thank the anonymous reviewers for their insightful comments and suggestions. This work is partially supported by the National Science Foundation under grant SCH-1838615, IIS-1553568, and a Google Faculty Research Award.
\end{acks}

\bibliographystyle{ACM-Reference-Format}
\bibliography{reference}


\begin{thebibliography}{34}


\ifx \showCODEN    \undefined \def \showCODEN     #1{\unskip}     \fi
\ifx \showDOI      \undefined \def \showDOI       #1{#1}\fi
\ifx \showISBNx    \undefined \def \showISBNx     #1{\unskip}     \fi
\ifx \showISBNxiii \undefined \def \showISBNxiii  #1{\unskip}     \fi
\ifx \showISSN     \undefined \def \showISSN      #1{\unskip}     \fi
\ifx \showLCCN     \undefined \def \showLCCN      #1{\unskip}     \fi
\ifx \shownote     \undefined \def \shownote      #1{#1}          \fi
\ifx \showarticletitle \undefined \def \showarticletitle #1{#1}   \fi
\ifx \showURL      \undefined \def \showURL       {\relax}        \fi
\providecommand\bibfield[2]{#2}
\providecommand\bibinfo[2]{#2}
\providecommand\natexlab[1]{#1}
\providecommand\showeprint[2][]{arXiv:#2}

\bibitem[\protect\citeauthoryear{Agarwal, Takatsu, Zaitsev, and
  Joachims}{Agarwal et~al\mbox{.}}{2019a}]%
        {Agarwal2019GFC}
\bibfield{author}{\bibinfo{person}{Aman Agarwal}, \bibinfo{person}{Kenta
  Takatsu}, \bibinfo{person}{Ivan Zaitsev}, {and} \bibinfo{person}{Thorsten
  Joachims}.} \bibinfo{year}{2019}\natexlab{a}.
\newblock \showarticletitle{A General Framework for Counterfactual
  Learning-to-Rank}. In \bibinfo{booktitle}{\emph{Proceedings of the 42Nd
  International ACM SIGIR Conference on Research and Development in Information
  Retrieval}} \emph{(\bibinfo{series}{SIGIR'19})}. \bibinfo{publisher}{ACM},
  \bibinfo{address}{New York, NY, USA}, \bibinfo{pages}{5--14}.
\newblock
\showISBNx{978-1-4503-6172-9}


\bibitem[\protect\citeauthoryear{Agarwal, Zaitsev, Wang, Li, Najork, and
  Joachims}{Agarwal et~al\mbox{.}}{2019b}]%
        {Agarwal2019EPB}
\bibfield{author}{\bibinfo{person}{Aman Agarwal}, \bibinfo{person}{Ivan
  Zaitsev}, \bibinfo{person}{Xuanhui Wang}, \bibinfo{person}{Cheng Li},
  \bibinfo{person}{Marc Najork}, {and} \bibinfo{person}{Thorsten Joachims}.}
  \bibinfo{year}{2019}\natexlab{b}.
\newblock \showarticletitle{Estimating Position Bias Without Intrusive
  Interventions}. In \bibinfo{booktitle}{\emph{Proceedings of the Twelfth ACM
  International Conference on Web Search and Data Mining}}
  \emph{(\bibinfo{series}{WSDM '19})}. \bibinfo{publisher}{ACM},
  \bibinfo{address}{New York, NY, USA}, \bibinfo{pages}{474--482}.
\newblock
\showISBNx{978-1-4503-5940-5}


\bibitem[\protect\citeauthoryear{Ai, Bi, Luo, Guo, and Croft}{Ai
  et~al\mbox{.}}{2018}]%
        {ai2018DLA}
\bibfield{author}{\bibinfo{person}{Qingyao Ai}, \bibinfo{person}{Keping Bi},
  \bibinfo{person}{Cheng Luo}, \bibinfo{person}{Jiafeng Guo}, {and}
  \bibinfo{person}{W.~Bruce Croft}.} \bibinfo{year}{2018}\natexlab{}.
\newblock \showarticletitle{Unbiased Learning to Rank with Unbiased Propensity
  Estimation}. In \bibinfo{booktitle}{\emph{The 41st International ACM SIGIR
  Conference on Research \&\#38; Development in Information Retrieval}}
  \emph{(\bibinfo{series}{SIGIR '18})}. \bibinfo{publisher}{ACM},
  \bibinfo{address}{New York, NY, USA}, \bibinfo{pages}{385--394}.
\newblock
\showISBNx{978-1-4503-5657-2}


\bibitem[\protect\citeauthoryear{Arias-Nicol{\'a}s, P{\'e}rez, and
  Mart{\'i}n}{Arias-Nicol{\'a}s et~al\mbox{.}}{2008}]%
        {Arias2008logistic}
\bibfield{author}{\bibinfo{person}{J.~P. Arias-Nicol{\'a}s},
  \bibinfo{person}{C.~J. P{\'e}rez}, {and} \bibinfo{person}{J. Mart{\'i}n}.}
  \bibinfo{year}{2008}\natexlab{}.
\newblock \showarticletitle{A logistic regression-based pairwise comparison
  method to aggregate preferences}.
\newblock \bibinfo{journal}{\emph{Group Decision and Negotiation}}
  \bibinfo{volume}{17}, \bibinfo{number}{3} (\bibinfo{date}{01 May}
  \bibinfo{year}{2008}), \bibinfo{pages}{237--247}.
\newblock
\showISSN{1572-9907}


\bibitem[\protect\citeauthoryear{Burges}{Burges}{2010}]%
        {burges2010from}
\bibfield{author}{\bibinfo{person}{Chris~J.C. Burges}.}
  \bibinfo{year}{2010}\natexlab{}.
\newblock \bibinfo{booktitle}{\emph{From RankNet to LambdaRank to LambdaMART:
  An Overview}}.
\newblock \bibinfo{type}{{T}echnical {R}eport} MSR-TR-2010-82.
\newblock


\bibitem[\protect\citeauthoryear{Burges, Ragno, and Le}{Burges
  et~al\mbox{.}}{2006}]%
        {Burges2006lambdarank}
\bibfield{author}{\bibinfo{person}{Christopher J.~C. Burges},
  \bibinfo{person}{Robert Ragno}, {and} \bibinfo{person}{Quoc~Viet Le}.}
  \bibinfo{year}{2006}\natexlab{}.
\newblock \showarticletitle{Learning to Rank with Nonsmooth Cost Functions}. In
  \bibinfo{booktitle}{\emph{Proceedings of the 19th International Conference on
  Neural Information Processing Systems}} \emph{(\bibinfo{series}{NIPS’06})}.
  \bibinfo{publisher}{MIT Press}, \bibinfo{address}{Cambridge, MA, USA},
  \bibinfo{pages}{193–200}.
\newblock


\bibitem[\protect\citeauthoryear{Chapelle, Joachims, Radlinski, and
  Yue}{Chapelle et~al\mbox{.}}{2012}]%
        {Chapelle2012interleave}
\bibfield{author}{\bibinfo{person}{Olivier Chapelle}, \bibinfo{person}{Thorsten
  Joachims}, \bibinfo{person}{Filip Radlinski}, {and} \bibinfo{person}{Yisong
  Yue}.} \bibinfo{year}{2012}\natexlab{}.
\newblock \showarticletitle{Large-Scale Validation and Analysis of Interleaved
  Search Evaluation}.
\newblock \bibinfo{journal}{\emph{ACM Trans. Inf. Syst.}} \bibinfo{volume}{30},
  \bibinfo{number}{1}, Article \bibinfo{articleno}{Article 6}
  (\bibinfo{date}{March} \bibinfo{year}{2012}), \bibinfo{numpages}{41}~pages.
\newblock
\showISSN{1046-8188}


\bibitem[\protect\citeauthoryear{Chapelle and Zhang}{Chapelle and
  Zhang}{2009}]%
        {Chapelle2009dynamicbayesian}
\bibfield{author}{\bibinfo{person}{Olivier Chapelle} {and} \bibinfo{person}{Ya
  Zhang}.} \bibinfo{year}{2009}\natexlab{}.
\newblock \showarticletitle{A Dynamic Bayesian Network Click Model for Web
  Search Ranking}. In \bibinfo{booktitle}{\emph{Proceedings of the 18th
  International Conference on World Wide Web}} \emph{(\bibinfo{series}{WWW
  ’09})}. \bibinfo{publisher}{ACM}, \bibinfo{pages}{1–10}.
\newblock
\showISBNx{9781605584874}


\bibitem[\protect\citeauthoryear{Chuklin, Markov, and de~Rijke}{Chuklin
  et~al\mbox{.}}{2016}]%
        {chuklin2015click}
\bibfield{author}{\bibinfo{person}{Aleksandr Chuklin}, \bibinfo{person}{Ilya
  Markov}, {and} \bibinfo{person}{Maarten de Rijke}.}
  \bibinfo{year}{2016}\natexlab{}.
\newblock \showarticletitle{Click Models for Web Search and Their Applications
  to IR: WSDM 2016 Tutorial}. In \bibinfo{booktitle}{\emph{Proceedings of the
  Ninth ACM International Conference on Web Search and Data Mining}}
  \emph{(\bibinfo{series}{WSDM ’16})}. \bibinfo{publisher}{ACM},
  \bibinfo{pages}{689–690}.
\newblock
\showISBNx{9781450337168}


\bibitem[\protect\citeauthoryear{Craswell, Zoeter, Taylor, and Ramsey}{Craswell
  et~al\mbox{.}}{2008}]%
        {Craswell2008clickpositionbias}
\bibfield{author}{\bibinfo{person}{Nick Craswell}, \bibinfo{person}{Onno
  Zoeter}, \bibinfo{person}{Michael Taylor}, {and} \bibinfo{person}{Bill
  Ramsey}.} \bibinfo{year}{2008}\natexlab{}.
\newblock \showarticletitle{An Experimental Comparison of Click Position-Bias
  Models}. In \bibinfo{booktitle}{\emph{Proceedings of the 2008 International
  Conference on Web Search and Data Mining}} \emph{(\bibinfo{series}{WSDM
  ’08})}. \bibinfo{publisher}{ACM}, \bibinfo{pages}{87–94}.
\newblock
\showISBNx{9781595939272}


\bibitem[\protect\citeauthoryear{Dupret and Piwowarski}{Dupret and
  Piwowarski}{2008}]%
        {Dupret2008UBM}
\bibfield{author}{\bibinfo{person}{Georges~E. Dupret} {and}
  \bibinfo{person}{Benjamin Piwowarski}.} \bibinfo{year}{2008}\natexlab{}.
\newblock \showarticletitle{A User Browsing Model to Predict Search Engine
  Click Data from Past Observations.}. In \bibinfo{booktitle}{\emph{Proceedings
  of the 31st Annual International ACM SIGIR Conference on Research and
  Development in Information Retrieval}} \emph{(\bibinfo{series}{SIGIR
  ’08})}. \bibinfo{publisher}{ACM}.
\newblock


\bibitem[\protect\citeauthoryear{Friedman}{Friedman}{2001}]%
        {Friedman2000mart}
\bibfield{author}{\bibinfo{person}{Jerome~H. Friedman}.}
  \bibinfo{year}{2001}\natexlab{}.
\newblock \showarticletitle{Greedy Function Approximation: A Gradient Boosting
  Machine}.
\newblock \bibinfo{journal}{\emph{The Annals of Statistics}}
  \bibinfo{volume}{29}, \bibinfo{number}{5} (\bibinfo{year}{2001}),
  \bibinfo{pages}{1189--1232}.
\newblock


\bibitem[\protect\citeauthoryear{Hu, Wang, Peng, and Li}{Hu
  et~al\mbox{.}}{2019}]%
        {hu2019unbiasedlambda}
\bibfield{author}{\bibinfo{person}{Ziniu Hu}, \bibinfo{person}{Yang Wang},
  \bibinfo{person}{Qu Peng}, {and} \bibinfo{person}{Hang Li}.}
  \bibinfo{year}{2019}\natexlab{}.
\newblock \showarticletitle{Unbiased LambdaMART: An Unbiased Pairwise
  Learning-to-Rank Algorithm}. In \bibinfo{booktitle}{\emph{The World Wide Web
  Conference}} \emph{(\bibinfo{series}{WWW '19})}. \bibinfo{publisher}{ACM},
  \bibinfo{address}{New York, NY, USA}, \bibinfo{pages}{2830--2836}.
\newblock
\showISBNx{978-1-4503-6674-8}


\bibitem[\protect\citeauthoryear{Imbens and Rubin}{Imbens and Rubin}{2015}]%
        {Imbens2015causal}
\bibfield{author}{\bibinfo{person}{Guido~W. Imbens} {and}
  \bibinfo{person}{Donald~B. Rubin}.} \bibinfo{year}{2015}\natexlab{}.
\newblock \bibinfo{booktitle}{\emph{Causal Inference for Statistics, Social,
  and Biomedical Sciences: An Introduction}}.
\newblock \bibinfo{publisher}{Cambridge University Press},
  \bibinfo{address}{USA}.
\newblock
\showISBNx{0521885884}


\bibitem[\protect\citeauthoryear{Jagerman, Oosterhuis, and de~Rijke}{Jagerman
  et~al\mbox{.}}{2019}]%
        {Jagerman2019modelorintervene}
\bibfield{author}{\bibinfo{person}{Rolf Jagerman}, \bibinfo{person}{Harrie
  Oosterhuis}, {and} \bibinfo{person}{Maarten de Rijke}.}
  \bibinfo{year}{2019}\natexlab{}.
\newblock \showarticletitle{To Model or to Intervene: A Comparison of
  Counterfactual and Online Learning to Rank from User Interactions}. In
  \bibinfo{booktitle}{\emph{Proceedings of the 42nd International ACM SIGIR
  Conference on Research and Development in Information Retrieval}}
  \emph{(\bibinfo{series}{SIGIR’19})}. \bibinfo{publisher}{ACM},
  \bibinfo{pages}{15–24}.
\newblock
\showISBNx{9781450361729}


\bibitem[\protect\citeauthoryear{Joachims}{Joachims}{2002}]%
        {Joachims2002OSE}
\bibfield{author}{\bibinfo{person}{Thorsten Joachims}.}
  \bibinfo{year}{2002}\natexlab{}.
\newblock \showarticletitle{Optimizing Search Engines Using Clickthrough Data}.
  In \bibinfo{booktitle}{\emph{Proceedings of the Eighth ACM SIGKDD
  International Conference on Knowledge Discovery and Data Mining}}
  \emph{(\bibinfo{series}{KDD '02})}. \bibinfo{publisher}{ACM},
  \bibinfo{address}{New York, NY, USA}, \bibinfo{pages}{133--142}.
\newblock
\showISBNx{1-58113-567-X}


\bibitem[\protect\citeauthoryear{Joachims, Granka, Pan, Hembrooke, and
  Gay}{Joachims et~al\mbox{.}}{2017a}]%
        {Joachims2017interpretclick}
\bibfield{author}{\bibinfo{person}{Thorsten Joachims}, \bibinfo{person}{Laura
  Granka}, \bibinfo{person}{Bing Pan}, \bibinfo{person}{Helene Hembrooke},
  {and} \bibinfo{person}{Geri Gay}.} \bibinfo{year}{2017}\natexlab{a}.
\newblock \showarticletitle{Accurately Interpreting Clickthrough Data as
  Implicit Feedback}.
\newblock \bibinfo{journal}{\emph{SIGIR Forum}} \bibinfo{volume}{51},
  \bibinfo{number}{1} (\bibinfo{date}{Aug.} \bibinfo{year}{2017}),
  \bibinfo{pages}{4–11}.
\newblock


\bibitem[\protect\citeauthoryear{Joachims, Granka, Pan, Hembrooke, Radlinski,
  and Gay}{Joachims et~al\mbox{.}}{2007}]%
        {Joachims2007Eval}
\bibfield{author}{\bibinfo{person}{Thorsten Joachims}, \bibinfo{person}{Laura
  Granka}, \bibinfo{person}{Bing Pan}, \bibinfo{person}{Helene Hembrooke},
  \bibinfo{person}{Filip Radlinski}, {and} \bibinfo{person}{Geri Gay}.}
  \bibinfo{year}{2007}\natexlab{}.
\newblock \showarticletitle{Evaluating the Accuracy of Implicit Feedback from
  Clicks and Query Reformulations in Web Search}.
\newblock \bibinfo{journal}{\emph{ACM Trans. Inf. Syst.}}
  (\bibinfo{date}{April} \bibinfo{year}{2007}).
\newblock


\bibitem[\protect\citeauthoryear{Joachims, Swaminathan, and Schnabel}{Joachims
  et~al\mbox{.}}{2017b}]%
        {joachims2017ips}
\bibfield{author}{\bibinfo{person}{Thorsten Joachims}, \bibinfo{person}{Adith
  Swaminathan}, {and} \bibinfo{person}{Tobias Schnabel}.}
  \bibinfo{year}{2017}\natexlab{b}.
\newblock \showarticletitle{Unbiased Learning-to-Rank with Biased Feedback}. In
  \bibinfo{booktitle}{\emph{Proceedings of the Tenth ACM International
  Conference on Web Search and Data Mining}} \emph{(\bibinfo{series}{WSDM
  '17})}. \bibinfo{publisher}{ACM}, \bibinfo{address}{New York, NY, USA},
  \bibinfo{pages}{781--789}.
\newblock
\showISBNx{978-1-4503-4675-7}


\bibitem[\protect\citeauthoryear{Qin, Chen, Metzler, Noh, Qin, and Wang}{Qin
  et~al\mbox{.}}{2020a}]%
        {attribute}
\bibfield{author}{\bibinfo{person}{Zhen Qin}, \bibinfo{person}{Suming~J. Chen},
  \bibinfo{person}{Donald Metzler}, \bibinfo{person}{Yongwoo Noh},
  \bibinfo{person}{Jingzheng Qin}, {and} \bibinfo{person}{Xuanhui Wang}.}
  \bibinfo{year}{2020}\natexlab{a}.
\newblock \showarticletitle{Attribute-Based Propensity for Unbiased Learning in
  Recommender Systems: Algorithm and Case Studies}. In
  \bibinfo{booktitle}{\emph{Proceedings of the 26th ACM SIGKDD International
  Conference on Knowledge Discovery and Data Mining}}
  \emph{(\bibinfo{series}{KDD '20})}. \bibinfo{pages}{2359–2367}.
\newblock


\bibitem[\protect\citeauthoryear{Qin, Li, Bendersky, and Metzler}{Qin
  et~al\mbox{.}}{2020b}]%
        {mcn}
\bibfield{author}{\bibinfo{person}{Zhen Qin}, \bibinfo{person}{Zhongliang Li},
  \bibinfo{person}{Michael Bendersky}, {and} \bibinfo{person}{Donald Metzler}.}
  \bibinfo{year}{2020}\natexlab{b}.
\newblock \showarticletitle{Matching Cross Network for Learning to Rank in
  Personal Search}. In \bibinfo{booktitle}{\emph{Proceedings of The Web
  Conference}} \emph{(\bibinfo{series}{WWW ’20})}.
  \bibinfo{pages}{2835–2841}.
\newblock


\bibitem[\protect\citeauthoryear{Rendle, Freudenthaler, Gantner, and
  Schmidt-Thieme}{Rendle et~al\mbox{.}}{2009}]%
        {rendle2009bpr}
\bibfield{author}{\bibinfo{person}{Steffen Rendle}, \bibinfo{person}{Christoph
  Freudenthaler}, \bibinfo{person}{Zeno Gantner}, {and} \bibinfo{person}{Lars
  Schmidt-Thieme}.} \bibinfo{year}{2009}\natexlab{}.
\newblock \showarticletitle{BPR: Bayesian Personalized Ranking from Implicit
  Feedback}. In \bibinfo{booktitle}{\emph{Proceedings of the Twenty-Fifth
  Conference on Uncertainty in Artificial Intelligence}}
  \emph{(\bibinfo{series}{UAI ’09})}. \bibinfo{publisher}{AUAI Press},
  \bibinfo{address}{Arlington, Virginia, USA}, \bibinfo{pages}{452–461}.
\newblock
\showISBNx{9780974903958}


\bibitem[\protect\citeauthoryear{Richardson, Dominowska, and Ragno}{Richardson
  et~al\mbox{.}}{2007}]%
        {richardson2007predicting}
\bibfield{author}{\bibinfo{person}{Matthew Richardson}, \bibinfo{person}{Ewa
  Dominowska}, {and} \bibinfo{person}{Robert Ragno}.}
  \bibinfo{year}{2007}\natexlab{}.
\newblock \showarticletitle{Predicting Clicks: Estimating the Click-through
  Rate for New Ads}. In \bibinfo{booktitle}{\emph{Proceedings of the 16th
  International Conference on World Wide Web}} \emph{(\bibinfo{series}{WWW
  ’07})}. \bibinfo{publisher}{ACM}, \bibinfo{pages}{521–530}.
\newblock
\showISBNx{9781595936547}


\bibitem[\protect\citeauthoryear{Rosenbaum and Rubin}{Rosenbaum and
  Rubin}{1983}]%
        {Paul1983central}
\bibfield{author}{\bibinfo{person}{Paul~R. Rosenbaum} {and}
  \bibinfo{person}{Donald~B. Rubin}.} \bibinfo{year}{1983}\natexlab{}.
\newblock \showarticletitle{The Central Role of the Propensity Score in
  Observational Studies for Causal Effects}.
\newblock \bibinfo{journal}{\emph{Biometrika}} \bibinfo{volume}{70},
  \bibinfo{number}{1} (\bibinfo{year}{1983}), \bibinfo{pages}{41--55}.
\newblock
\showISSN{00063444}


\bibitem[\protect\citeauthoryear{Schnabel, Swaminathan, Singh, Chandak, and
  Joachims}{Schnabel et~al\mbox{.}}{2016}]%
        {Schnabel2016RecTreat}
\bibfield{author}{\bibinfo{person}{Tobias Schnabel}, \bibinfo{person}{Adith
  Swaminathan}, \bibinfo{person}{Ashudeep Singh}, \bibinfo{person}{Navin
  Chandak}, {and} \bibinfo{person}{Thorsten Joachims}.}
  \bibinfo{year}{2016}\natexlab{}.
\newblock \showarticletitle{Recommendations As Treatments: Debiasing Learning
  and Evaluation}. In \bibinfo{booktitle}{\emph{Proceedings of the 33rd
  International Conference on International Conference on Machine Learning -
  Volume 48}} \emph{(\bibinfo{series}{ICML'16})}.
  \bibinfo{publisher}{JMLR.org}, \bibinfo{pages}{1670--1679}.
\newblock


\bibitem[\protect\citeauthoryear{Swaminathan and Joachims}{Swaminathan and
  Joachims}{2015a}]%
        {Swaminathan2015batchlearning}
\bibfield{author}{\bibinfo{person}{Adith Swaminathan} {and}
  \bibinfo{person}{Thorsten Joachims}.} \bibinfo{year}{2015}\natexlab{a}.
\newblock \showarticletitle{Batch Learning from Logged Bandit Feedback through
  Counterfactual Risk Minimization}.
\newblock \bibinfo{journal}{\emph{Journal of Machine Learning Research}}
  \bibinfo{volume}{16}, \bibinfo{number}{52} (\bibinfo{year}{2015}).
\newblock


\bibitem[\protect\citeauthoryear{Swaminathan and Joachims}{Swaminathan and
  Joachims}{2015b}]%
        {adith2015selfnormalized}
\bibfield{author}{\bibinfo{person}{Adith Swaminathan} {and}
  \bibinfo{person}{Thorsten Joachims}.} \bibinfo{year}{2015}\natexlab{b}.
\newblock \showarticletitle{The Self-Normalized Estimator for Counterfactual
  Learning}.
\newblock In \bibinfo{booktitle}{\emph{Advances in Neural Information
  Processing Systems 28}}. \bibinfo{publisher}{Curran Associates, Inc.},
  \bibinfo{pages}{3231--3239}.
\newblock


\bibitem[\protect\citeauthoryear{Wang, Zhai, Dong, and Chang}{Wang
  et~al\mbox{.}}{2013}]%
        {Wang2013contentaware}
\bibfield{author}{\bibinfo{person}{Hongning Wang}, \bibinfo{person}{ChengXiang
  Zhai}, \bibinfo{person}{Anlei Dong}, {and} \bibinfo{person}{Yi Chang}.}
  \bibinfo{year}{2013}\natexlab{}.
\newblock \showarticletitle{Content-Aware Click Modeling}. In
  \bibinfo{booktitle}{\emph{Proceedings of the 22nd International Conference on
  World Wide Web}} \emph{(\bibinfo{series}{WWW ’13})}.
  \bibinfo{publisher}{ACM}, \bibinfo{pages}{1365–1376}.
\newblock
\showISBNx{9781450320351}


\bibitem[\protect\citeauthoryear{Wang, Bendersky, Metzler, and Najork}{Wang
  et~al\mbox{.}}{2016}]%
        {Wangselectionbias}
\bibfield{author}{\bibinfo{person}{Xuanhui Wang}, \bibinfo{person}{Michael
  Bendersky}, \bibinfo{person}{Donald Metzler}, {and} \bibinfo{person}{Marc
  Najork}.} \bibinfo{year}{2016}\natexlab{}.
\newblock \showarticletitle{Learning to Rank with Selection Bias in Personal
  Search}. In \bibinfo{booktitle}{\emph{Proceedings of the 39th International
  ACM SIGIR Conference on Research and Development in Information Retrieval}}
  \emph{(\bibinfo{series}{SIGIR ’16})}. \bibinfo{publisher}{ACM}.
\newblock
\showISBNx{9781450340694}


\bibitem[\protect\citeauthoryear{Wang, Golbandi, Bendersky, Metzler, and
  Najork}{Wang et~al\mbox{.}}{2018a}]%
        {wang2018pbe}
\bibfield{author}{\bibinfo{person}{Xuanhui Wang}, \bibinfo{person}{Nadav
  Golbandi}, \bibinfo{person}{Michael Bendersky}, \bibinfo{person}{Donald
  Metzler}, {and} \bibinfo{person}{Marc Najork}.}
  \bibinfo{year}{2018}\natexlab{a}.
\newblock \showarticletitle{Position Bias Estimation for Unbiased Learning to
  Rank in Personal Search}. In \bibinfo{booktitle}{\emph{Proceedings of the
  Eleventh ACM International Conference on Web Search and Data Mining}}
  \emph{(\bibinfo{series}{WSDM '18})}. \bibinfo{publisher}{ACM},
  \bibinfo{address}{New York, NY, USA}, \bibinfo{pages}{610--618}.
\newblock
\showISBNx{978-1-4503-5581-0}


\bibitem[\protect\citeauthoryear{Wang, Li, Golbandi, Bendersky, and
  Najork}{Wang et~al\mbox{.}}{2018b}]%
        {Wang2018Lambdaloss}
\bibfield{author}{\bibinfo{person}{Xuanhui Wang}, \bibinfo{person}{Cheng Li},
  \bibinfo{person}{Nadav Golbandi}, \bibinfo{person}{Michael Bendersky}, {and}
  \bibinfo{person}{Marc Najork}.} \bibinfo{year}{2018}\natexlab{b}.
\newblock \showarticletitle{The LambdaLoss Framework for Ranking Metric
  Optimization}. In \bibinfo{booktitle}{\emph{Proceedings of the 27th ACM
  International Conference on Information and Knowledge Management}}
  \emph{(\bibinfo{series}{CIKM '18})}. \bibinfo{publisher}{ACM},
  \bibinfo{address}{New York, NY, USA}, \bibinfo{pages}{1313--1322}.
\newblock
\showISBNx{978-1-4503-6014-2}


\bibitem[\protect\citeauthoryear{Yue and Joachims}{Yue and Joachims}{2009}]%
        {yue2009interactively}
\bibfield{author}{\bibinfo{person}{Yisong Yue} {and} \bibinfo{person}{Thorsten
  Joachims}.} \bibinfo{year}{2009}\natexlab{}.
\newblock \showarticletitle{Interactively Optimizing Information Retrieval
  Systems as a Dueling Bandits Problem}. In
  \bibinfo{booktitle}{\emph{Proceedings of the 26th Annual International
  Conference on Machine Learning}} \emph{(\bibinfo{series}{ICML ’09})}.
  \bibinfo{publisher}{ACM}, \bibinfo{pages}{1201–1208}.
\newblock
\showISBNx{9781605585161}


\bibitem[\protect\citeauthoryear{Yue, Patel, and Roehrig}{Yue
  et~al\mbox{.}}{2010}]%
        {yue2010beyondposition}
\bibfield{author}{\bibinfo{person}{Yisong Yue}, \bibinfo{person}{Rajan Patel},
  {and} \bibinfo{person}{Hein Roehrig}.} \bibinfo{year}{2010}\natexlab{}.
\newblock \showarticletitle{Beyond Position Bias: Examining Result
  Attractiveness as a Source of Presentation Bias in Clickthrough Data}. In
  \bibinfo{booktitle}{\emph{Proceedings of the 19th International Conference on
  World Wide Web}} \emph{(\bibinfo{series}{WWW ’10})}.
  \bibinfo{publisher}{ACM}, \bibinfo{pages}{1011–1018}.
\newblock


\bibitem[\protect\citeauthoryear{Zhu and Klabjan}{Zhu and Klabjan}{2020}]%
        {zhu2020listwise}
\bibfield{author}{\bibinfo{person}{Xiaofeng Zhu} {and} \bibinfo{person}{Diego
  Klabjan}.} \bibinfo{year}{2020}\natexlab{}.
\newblock \showarticletitle{Listwise Learning to Rank by Exploring Unique
  Ratings}. In \bibinfo{booktitle}{\emph{Proceedings of the 13th International
  Conference on Web Search and Data Mining}} \emph{(\bibinfo{series}{WSDM
  ’20})}. \bibinfo{publisher}{ACM}, \bibinfo{pages}{798–806}.
\newblock
\showISBNx{9781450368223}


\end{thebibliography}

\clearpage
\onecolumn
\appendix

\end{document}